\documentclass{aa}

\usepackage{graphicx}

\usepackage{txfonts}

\usepackage{natbib}

\begin{document}

   \title{Formation Dynamics of a Rotating Solar Spicule-like Event and Associated Energy-Mass Transport}

    \author{
        Liufan Gong\inst{1,2}
        \and
        Lei Ni\inst{1,2,3}\thanks{Corresponding author: \email{leini@ynao.ac.cn}}
        \and
        Guanchong Cheng\inst{1,2,3}\thanks{Corresponding author: \email{chengguanchong@ynao.ac.cn}}
        \and
        Jiajia Liu\inst{4,5}
        \and
        Quan Xie\inst{4,5}
        \and
        Robert Erd\'elyi\inst{6,7,8}
        }

   \institute{Yunnan Observatories, Chinese Academy of Sciences, Kunming, Yunnan 650216, PR China
              \and
              University of Chinese Academy of Sciences, Beijing 100049, PR China
              \and
              Yunnan Key Laboratory of Solar Physics and Space Science, Kunming 650216, PR China
              \and
              National Key Laboratory of Deep Space Exploration, School of Earth and Space Sciences, University of Science and Technology of China, Hefei 230026, China
              \and
              CAS Center for Excellence in Comparative Planetology/CAS Key Laboratory of Geospace Environment/Mengcheng National Geophysical Observatory, University of Science and Technology of China, Hefei 230026, China
              \and
              Solar Physics and Space Plasma Research Centre, School of Mathematical and Physical Sciences, University of Sheffield, Hicks Building, Hounsfield Road, Sheffield S3 7RH, United Kingdom
              \and
              Department of Astronomy, E{\"o}tv{\"o}s Lor{\'a}nd University, P{\'a}zm{\'a}ny P{\'e}ter s{\'e}t{\'a}ny 1/A, Budapest, H-1112, Hungary
              \and
              Gyula Bay Zolt{\'a}n Solar Observatory (GSO), Hungarian Solar Physics Foundation (HSPF), Pet\H{o}fi t{\'e}r 3, Gyula, H-5700, Hungary}

\abstract
{Vortical structures and elongated chromospheric spicule-like features are ubiquitous small-scale dynamic phenomena in the lower solar atmosphere. Both are important channels for transporting mass and energy from the solar surface to the corona. However, the physical connection between vortical motions and the formation of spicule-like structures, as well as their role in mass and energy transport, remains incompletely understood.}
{We aim to investigate  the connection between the vortex motion and the formation of a spicule-like event and the associated mass and energy transport.}
{We performed three-dimensional radiative magnetohydrodynamic(3D RMHD) simulations in an upward-directed (positive-polarity) open magnetic field with an initial background strength of 10~G. We identified vortex-associated spicule-like events and carried out a detailed analysis of one representative event. }
{Prior to the formation of this spicule-like event, a photospheric vortex appears and twists the magnetic field lines, then develops upward into higher atmospheric layers, forming a magnetic tornado system. This event is generated during this process. At the periphery of this magnetic tornado, twisted magnetic field lines with negative vertical components reconnect with both external positive-polarity background fields and central positive-polarity fields. The magnetic reconnection sites rise and rotate together with the developed magnetic tornado, and many jet-like substructures in this event are closely associated with plasma upward ejections from these reconnection regions. A vertical force analysis shows that, for this event, plasma uplift in the photosphere and lower chromosphere is dominated by the pressure-gradient force, whereas in the upper chromosphere, the Lorentz force gradually becomes dominant. At a height of about $4$\,Mm above the solar surface, this event produces a mean upward mass flux density on the order of $10^{-7}~{\rm kg~m^{-2}~s^{-1}}$ and a mean upward Poynting flux density of approximately $525~{\rm W~m^{-2}}$. The wave-mode decomposition results show that the energy transport associated with the rotational spicule-like event is dominated by the Alfv\'en mode. Alfv\'en waves, fast-mode waves, and slow-mode waves carry energy into the lower corona in an approximate ratio of $10:7:3$.}
{Our results demonstrate that spicule-like events can be driven jointly by the vortex motion in the lower atmosphere and the induced magnetic reconnection. During its formation process, considerable mass and energy are transported to the lower corona.}


   \maketitle

\section{Introduction}
  
High-resolution solar observations and advanced numerical simulations have shown that vortex motions (also called swirls) are widespread in the lower solar atmosphere, occurring in both quiet-Sun and active regions \citep[e.g.,][]{Attie2009,Balmaceda2010,Tziotziou2023}. In the photosphere, they manifest as small-scale rotational flow structures in intergranular lanes \citep[e.g.,][]{Brandt1988,Bonet2008}. Statistical studies further show that the spatial scales of photospheric swirls range from hundreds to thousands of kilometres, with lifetimes from several minutes to several hours, and an average rotational velocity of about hundreds of metres per second to several kilometres per second. \citep[e.g.,][]{Vargas2011,Giagkiozis2018,Liu2019b,Xie2025}. In the chromosphere, vortex motions are often manifested as arc-shaped, spiral-arm-like, or ring-shaped structures, with typical diameters of about 1.5--2 Mm and associated outward flows of 2--4\,km\,s$^{-1}$ \citep[e.g.,][]{Wedemeyer2009,Park2016,Shetye2019,Dakanalis2022}. In the corona, the observed rotation of coronal loops and the twisted structures at coronal hole boundaries are also confirmed to be closely associated with vortex motions, and particularly noteworthy is the coherent rotating magnetic field structure induced by vortices — i.e., magnetic tornadoes \citep[e.g.,][]{Pike1998,Li2012,Liu2012,Su2012,Wedemeyer2012,Wedemeyer2013,Panesar2013,Wedemeyer2014,Wang2017,Tziotziou2018}. By perturbing both the plasma and the magnetic field, vortex motions may provide an efficient channel for the upward transport of mass, momentum, and energy from the lower solar atmosphere \citep[e.g.,][]{Shelyag2011,Wedemeyer2012,Murawski2018,Liu2019NatCo,Yadav2020,Scalisi2021,Scalisi2023,Breu2023,Tziotziou2023}.

Numerical studies have shown that photospheric vortex or spiral motions can excite MHD waves and transport wave energy upward through the solar atmosphere \citep[e.g.,][]{Mumford2015,Mumford2015MN,Leonard2018}. More recently, \citet{Finley2022} found in their simulation that photospheric swirls can twist the coronal magnetic field and excite torsional Alfv\'en waves, yielding a local maximum Poynting flux density of  $2$--$4 \times 10^{3}$\,W\,m$^{-2}$ in the low corona. Meanwhile, nonlinear interactions between vortices can lead to the formation of small-scale current sheets and shocks, thereby converting kinetic and magnetic energy into thermal energy. \citet{Battaglia2021} investigated the origin and propagation of typical chromospheric vortices in observations and simulations, and identified unidirectional vortices propagating from magnetic footpoints into the chromosphere at approximately the Alfv\'en speed. The RMHD simulations also proved that the current sheets with strong currents can be formed between the interacting vortices \citep[e.g.,][]{Yadav2021}, and the dissipation of these current sheets contributes to the heating of chromosphere. \citet{Kuniyoshi2023} further showed that the coalescence of magnetic flux concentrations can trigger magnetic tornadoes. In the presence of magnetic tornadoes, the time-averaged Poynting flux measured at the transition region was found to be about four times larger than in cases without magnetic tornadoes. These studies suggest that vortex motions provide an efficient channel for the transport and deposition of energy from the convection zone to the chromosphere and corona. In particular, torsional Alfv\'en waves excited by vortex tubes can carry a substantial amount of energy upward \citep[e.g.,][]{Liu2019NatCo,Oxley2020,Scalisi2021}
  
Solar spicules \cite[e.g.,][]{2000SoPh..196...79S, Pereira2012,2012SSRv..169..181T} are small-scale, jet-like features \citep[e.g.,][]{2014Sci...346A.315T, Chen2017, 2019STP.....5b..58H, 2021RSPSA.47700217S, 2022A&A...657A.118Q} with widths of about 1 Mm or less. They are elongated, dynamic, cool plasma structures ejected from the lower atmosphere, with typical lengths of 2--10 Mm \citep[e.g.,][]{Beckers1963,Sterling2000,Zaqarashvili2009,Tsiropoula2012}. Although extensive observational and theoretical studies have been conducted on spicules, the primary physical mechanisms driving the generation of spicules in different regions are still not fully understood. A variety of mechanisms have been proposed, including upward-propagating acoustic or magnetoacoustic shocks and related wave-driven processes \citep[e.g.,][]{Shibata1982,Iijima2015,Dey2022,Srivastava2025}, Alfv\'en waves or upward Lorentz forces generated by twisted magnetic field lines \citep[e.g.,][]{Hollweg1982,DePontieu1998,James2002}, magnetic reconnection outflows \citep[e.g.,][]{Yokoyama1995,Singh2011,Archontis2010,Yang2013}, and enhanced Lorentz forces associated with ambipolar diffusion \citep[e.g.,][]{Mart2017}. Given the strong stratification, rapid evolution, and complex thermodynamics of the chromosphere, spicules are often not driven by a single mechanism but are likely the result of a combination of multiple driving mechanisms\citep{Li2026}.

A number of observational studies have reported that many spicules exhibit rotational motions during their formation and evolution \citep[e.g.,][]{Zaqarashvili2009,DePontieu2012,DePontieu2014}, suggesting that spicule formation may be related to vortex motions in the lower solar atmosphere. Recent numerical studies have also indicated that photospheric and chromospheric vortices may be closely associated with jet formation. For example, \citet{Kitiashvili2013} found that turbulent vortex tubes can trigger large-scale jets, which are mainly accelerated by pressure gradients below and near the photosphere, and by Lorentz forces at higher atmospheric layers. The numerical simulation in \citet{Iijima2017} showed that a chromospheric jet formation originated from vortex motions, and the resulted upward Lorentz force was found as the main driving force. \citet{Snow2018} investigated the interaction of vortical motions in a pair of expanding and merging magnetic flux tubes, showing that such interactions can rearrange the magnetic field into a multi-threaded structure and drive high-speed shocks propagating into the upper solar atmosphere at about 50 km s$^{-1}$. This provides a new mechanism for driving chromospheric spicules through the interaction of vortex motions in merging magnetic flux tubes. More recently, \citet{Dey2026} found persistent vortex tubes extending into the corona in synthetic spicule clusters.These results suggest that vortical flows not only guide wave propagation, but may also drive jet-like upflows and shocks, thereby efficiently transporting mass to higher atmospheric layers \citep[e.g.,][]{Ni2026}.

To further investigate the connection between vortices and spicules, we performed a high-resolution 3D RMHD simulation by using the MURaM code. The simulation reveals that numerous spicule formation processes are associated with the vortex motions in the lower atmosphere. In this paper, we focus on one representative vortex-driven spicule event and study the underlying physical processes in detail, together with the associated mass and energy transport. We find that the formation of the spicule results from the combined action of multiple driving mechanisms. Through the wave-mode decomposition method, we quantify the relative contributions of different MHD wave modes for carrying energy upward during this process. The remaining part of this paper is organised as follows. Section~2 describes the numerical setup and methods. Section~3 presents the simulation results. We present the conclusions and discussions in Section~4.
   
\section{Numerical methods and setup}

\begin{figure*}
\sidecaption
\includegraphics[width=12cm]{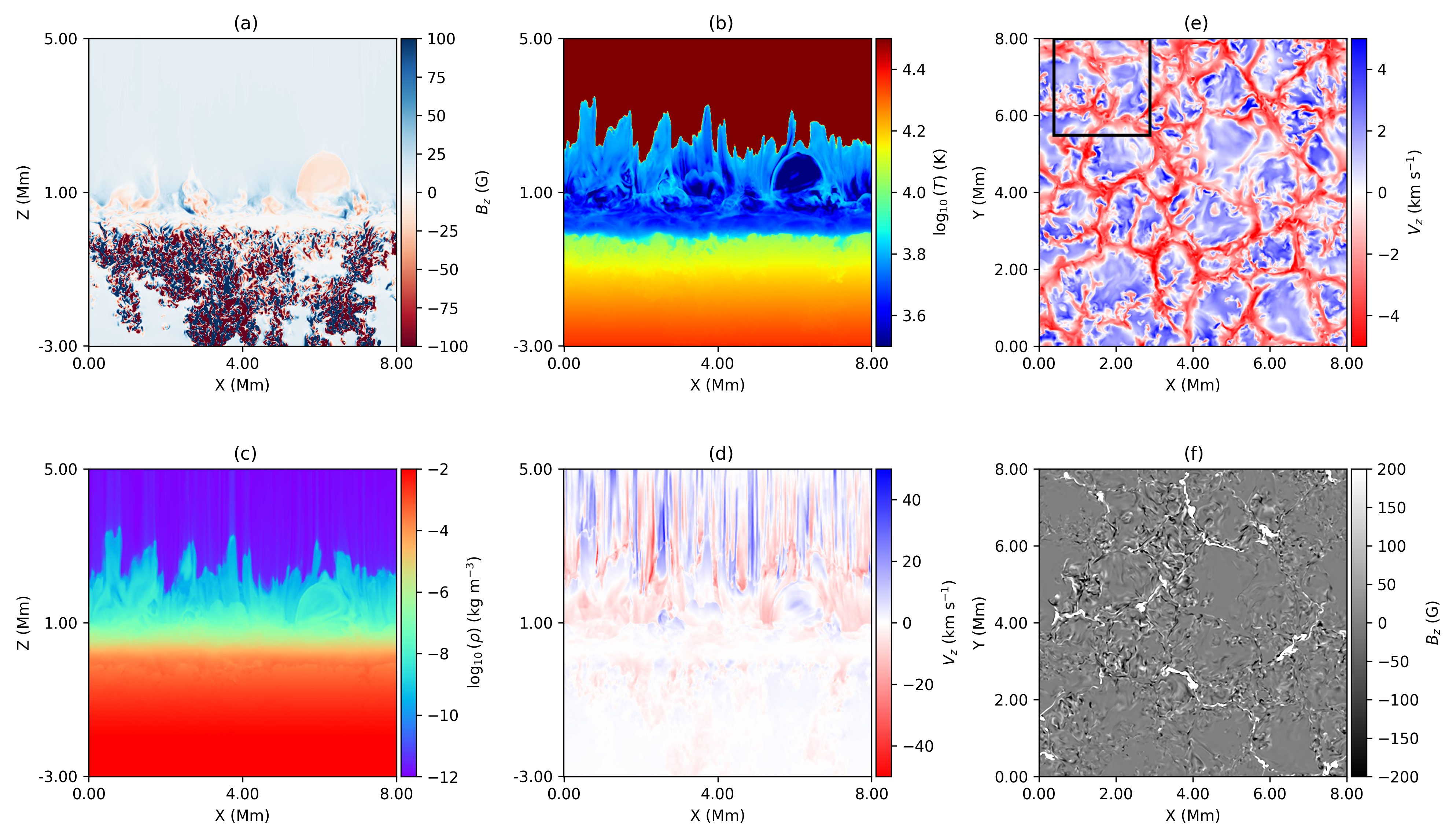}
\caption{Overview of the simulation domain at $t=2832$~s. Panels (a)--(d) show the distributions of  vertical magnetic field ($B_z$), logarithm of temperature ($log_{10} (T)$), logarithm of density ($log_{10} (\rho)$), and vertical velocity ($V_z$) in an $x-z$ vertical slice, respectively. Panel (e) shows the vertical velocity distribution ($V_z$) in the $x-y$ plane at $z=0$\,Mm, and the black rectangular box marks the region selected for detailed analysis in the following figures. Panel (f) shows the distribution of vertical magnetic field ($B_z$) in the $x-y$ plane at $z=0$\,Mm.}
\label{fig:overview}
\end{figure*}

The simulation is performed with an extended version of the MURaM radiative magnetohydrodynamic(RMHD) code, based on the implementations described by \citet{Rempel2017} and \citet{Przybylski2022}. The model extends from the upper convection zone into the corona. Radiative transfer in the photosphere and lower chromosphere is treated in LTE using a grey opacity table, while the parametrised non-LTE treatment \citep{Przybylski2022} is performed in the chromosphere, including H$\alpha$, Mg~\textsc{ii} h and k, and Ca~\textsc{ii} H and K lines. In the corona, optically thin radiative losses and field-aligned Spitzer-type thermal conduction are included. The coronal back-heating term and ambipolar diffusion are disabled in this simulation. Thermodynamic quantities are obtained from the merged Uppsala equation-of-state table described by \citet{Gudiksen2011}. Similar versions of MURaM have been used to study a range of small- and intermediate-scale magnetic phenomena in the lower solar atmosphere, including turbulent magnetic reconnection \citep[e.g.,][]{Cheng2025}, coronal jets \citep[e.g.,][]{Chen2026}, and solar spicules \citep[e.g.,][]{Chandra2025}.

The computational domain covers $8~{\rm Mm}\times8~{\rm Mm}\times8~{\rm Mm}$. The $z$-axis is perpendicular to the solar surface, and the domain extends from $z=-3$~Mm in the upper convection zone to $z=5$~Mm in the corona, with $z=0$ corresponding to the photosphere. A uniform grid was used in all three directions,

$$
    \Delta x=\Delta y=\Delta z=15.625~{\rm km},
$$
giving a total resolution of $512^{3}$ grid cells.

The initial atmosphere is constructed specifically for this simulation rather than taken from a previously relaxed hydrodynamic model. In the convection zone, we use the standard solar model to set the initial plasma parameters. Beyond the height range covered by the standard solar model, the initial density, pressure, and temperature profiles are obtained from interpolations by assuming isothermal hydrostatic stratification with a constant pressure scale height evaluated at the matching point. Pressure and density are extrapolated exponentially with height, while the temperature in the atmosphere is set to the value at the top of the standard solar model. The corresponding specific internal energy is then obtained from the Uppsala equation-of-state table using the local density and temperature. The resulting one-dimensional stratification is replicated in the horizontal directions.

No fixed temperature floor is imposed. Instead, the specific internal energy is prevented from falling below $1.75 \times 10^{12}$\,erg\,g$^{-1}$. Because temperature is obtained from density and specific internal energy through the tabulated equation of state, there is no fixed minimum temperature throughout the domain. If radiative cooling is extremely strong in some localised regions, it can cause negative energy to appear within just one iteration of an MHD time step. In that case, the magnitude of radiative cooling is artificially reduced, which is reflected in the code as a comparison between two integration time steps. In the low-density upper atmosphere, the time-step constraint imposed by large Alfv\'en speeds was alleviated using the semi-relativistic Boris correction \citep{Rempel2017}. The reduced characteristic speed was adjusted dynamically according to
\begin{equation}
c_{\rm red}=\max\left(100~{\rm km s^{-1}}, 3v_{\max}, 2c_{s,\max}\right),
\end{equation}
where $v_{\max}$ and $c_{s,\max}$ denote the maximum flow speed and the sound speed entering the timestep control, respectively. The Boris correction leaves stationary force balance unchanged but modifies wave propagation where the physical Alfv\'en speed approaches $c_{\rm red}$. Its effects are expected to be most relevant in strongly magnetised, low-density regions of the upper atmosphere.

The physical viscosity and resistivity are not included in this work. The dissipation is supplied by the local, nonlinear, slope-limited numerical diffusion in MURaM \citep{Rempel2014,Rempel2017,Przybylski2022}. This numerical diffusion depends on both the grid size and local solution, even the use of an isotropic numerical grid will not give isotropic diffusivities. Therefore, the numerical diffusion is not uniform in this work. The numerical resistive-heating diagnostic $Q_{\rm res}$ and viscous-heating diagnostic $Q_{\rm vis}$ are retained and can be used to calculate the numerical diffusivity and viscosity. The magnitude of the measured numerical diffusivity is roughly at the order of $10^{10}$\,cm$^2$\,s$^{-1}$, and the averaged magnetic Prandtl number in the convection zone is about $0.5$. The smallest resolved structures, including current concentration regions, are characterised jointly by the $15.625$\,km grid spacing and the local numerical diffusion. The mean plasma $\beta$ decreases with height; the region where the mean plasma $\beta$ is unity generally lies between $z = 1.1$\,Mm and $z = 1.4$\,Mm.

Periodic boundary conditions are applied in the horizontal directions. At the lower boundary, an isentropic condition allows both inflows and outflows, while the magnetic field follows the open symmetric field boundary condition (OSb) introduced by \citet{Rempel2014}. At the top, the flow boundary is half open: outflows are allowed and inflows are suppressed. The magnetic field is extrapolated as a potential field. The hot-plate condition is active throughout the run and we fix the specific internal energy at the upper boundary to $2 \times 10^{14}$\,erg\,g$^{-1}$. It therefore acts as an upper-boundary condition used to maintain the hot upper atmosphere; it is not a volume-distributed coronal heating term. The numerical boundary treatment covers $80$\,km, or about five grid cells. We retain the standard boundary diffusivity, with unit velocity and magnetic diffusion multipliers and no additional slope reduction at the up boundary. Box-mode damping is also used to suppress the global vertical oscillation of the domain. 

The semi-open top boundary is expected to produce weaker reflections than a closed boundary, but it is not a strictly reflection-free characteristic boundary. Nevertheless, we did not observe clear evidence that reflected disturbances dominate the evolution of the analysed event. Furthermore, our analysis is confined to heights below $4.5$\,Mm, which is well separated from the top boundary at $z = 5$\,Mm. The swirl height may be limited by the vertical extent of our simulation box \citep{Cherry2026}; this possibility should be investigated in the future by using an identical setup with a taller box.

At $t=0$, the atmosphere contains no developed convective flow. Convection is seeded with independent random perturbations in all three velocity components, distributed between $-100$\,cm\,s$^{-1}$ and $100$\,cm\,s$^{-1}$. A uniform vertical magnetic field of $10$\,G is initially imposed. We analysed the data after $2800$\,s in this work. By then, the convection is well developed and the initially uniform field has been re-organised into a strongly structured, mixed-polarity small-scale field. The distribution of vertical magnetic field ($B_z$) at the solar surface (see Figure~\ref{fig:overview}(f) ) is similar as the result for the saturated phase shown in the previous work \citep{Rempel2014}. However, we did not carry out the statistical analysis needed to establish saturation of the small-scale dynamo. During the interval of interest, full three-dimensional snapshots were stored every $2$\,s.

Figure~\ref{fig:overview} gives an overview of the simulation results at $t=2832$\,s. Figures~\ref{fig:overview}(a)--(d) show the distributions of  vertical magnetic field, logarithm of temperature, logarithm of density and vertical velocity, respectively, in an $x$--$z$ cross section. The temperature and density distributions illustrate the strong atmospheric stratification across the domain. Numerous spicule-like events are already present at this stage, and inspection of their evolution shows that many are associated with vortex motions in the lower atmosphere. Figure~\ref{fig:overview}(e) shows the vertical velocity distribution at the photosphere ($z=0$\,Mm); the black rectangular box marks the region selected for detailed analyses of the representative event in the following section.

\section{Simulation results}
\subsection{A Spicule-like event associated with vortex motion}

\begin{figure*}
\centering
\includegraphics[width=0.8\textwidth,keepaspectratio]{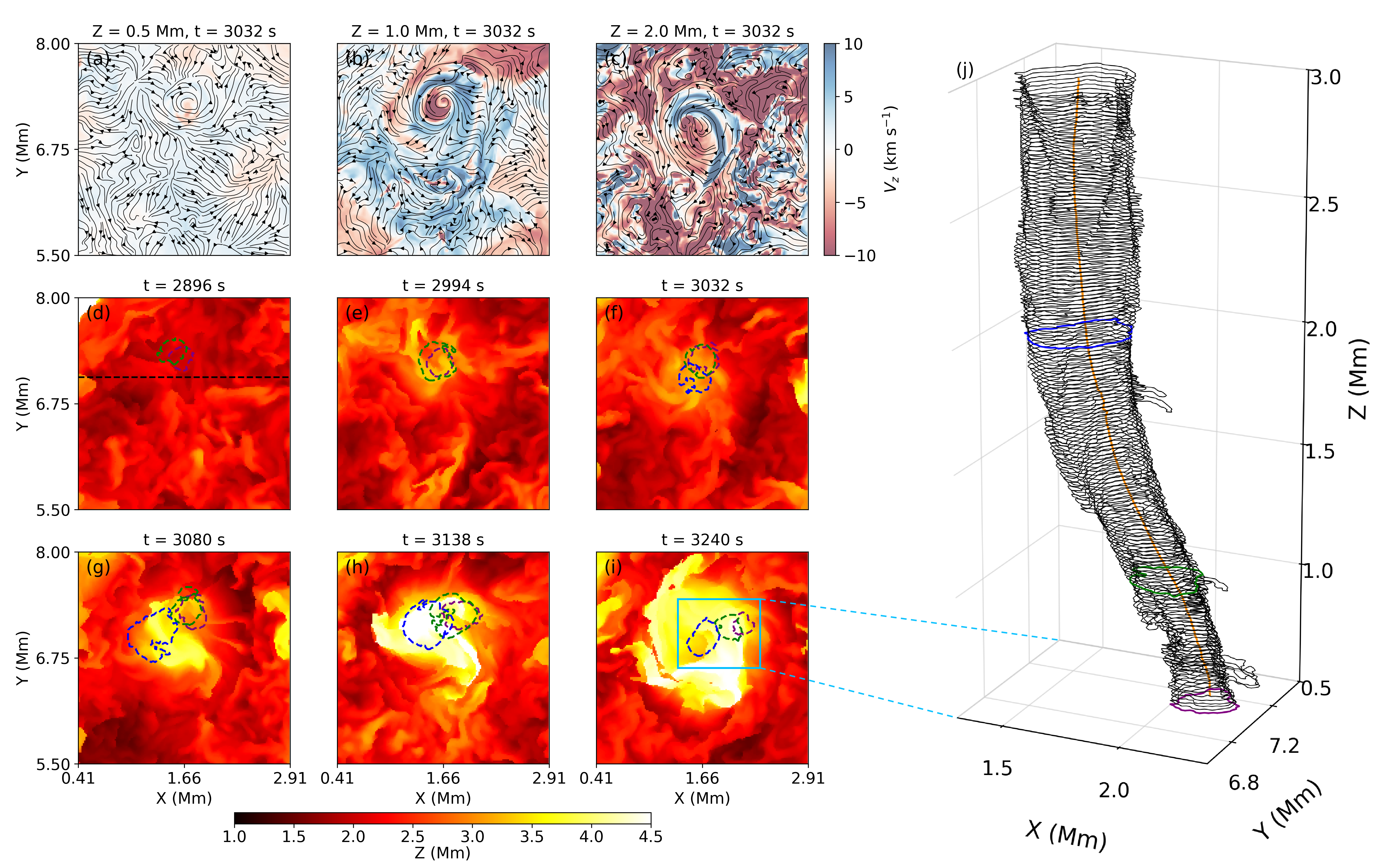}
\caption{Evolution of the vortex structure and the associated spicule. 
Panels (a)--(c) show the vertical velocity distributions at $z=0.5$\,Mm, $1$\,Mm, and $2$\,Mm at $t=3032$\,s, respectively. The black curves with arrows indicate the velocity streamlines. 
Panels (d)--(i) show the transition-region height distributions at six representative times. The black dashed line in panel (d) marks the fixed locations used for the subsequent analyses in the $x$--$z$ plane.
The purple, green, and blue dashed irregular circles  denote the identified vortex boundaries at $z=0.5$\,Mm, $1$\,Mm, and $2$\,Mm, respectively. Panel (j) shows the vortex boundaries at different heights, and the orange line in the centre indicates the axis of the vortex system from the lower atmosphere to the corona. The associated movie shows the temporal evolution of the transition-region height.
The associated movie is available online.} 
\label{fig:vortex_spicule_evolution}
\end{figure*}

Figure~\ref{fig:vortex_spicule_evolution} illustrates the temporal evolution of the vortex morphology and the associated spicule in the selected region. The velocity streamlines in Figures~\ref{fig:vortex_spicule_evolution}(a)--(c) show that, at $t=3032$~s, clear vortex motions are present in the horizontal $x$--$y$ plane with velocity field map at heights of $z=0.5$\,Mm, $1$\,Mm, and $2$\,Mm. The vortex radius becomes larger with increasing height, while the vertical upward flows are mainly concentrated in the peripheral region of the vortex. 

Figure~\ref{fig:vortex_spicule_evolution}(d)--(i) present six representative snapshots of the event evolution. Here, the spicule structure is represented by a transition-region height map: for each $(x,y)$ column, the height at which the temperature first reaches $4\times10^{4}$~K is identified and colour-coded. This method provides an intuitive way to display the overall morphology of the spicule top and the upward displacement of the transition region \citep{Iijima2017}. Figure~\ref{fig:vortex_spicule_evolution}(d) corresponds to the early stage of the event, when clear vortices have already formed at $0.5$\,Mm and $1$\,Mm, but no obvious spicule structure is yet visible. Figure~\ref{fig:vortex_spicule_evolution}(e)--(f) begin to show signatures of a local uplift of the transition region. Part of the spicule structure can already be identified, although its overall height is still relatively low. Figure~\ref{fig:vortex_spicule_evolution}(g)--(i) correspond to a more developed stage, during which the spicule structure becomes clearly visible.

The identified vortex boundaries at different heights are overlaid on these panels from (d) to (i), where the purple, green, and blue dashed contours denote the vortex boundaries at $z = 0.5$\,Mm, $1$\,Mm and $2$\,Mm, respectively. It should be noted that, in the early stage shown in Figure~\ref{fig:vortex_spicule_evolution}(d)--(e), vortices can only be identified at $z = 0.5$ and $1$\,Mm (purple and green dashed contours). At the subsequent stages, the vortex locations at all three heights show a good spatial correspondence with the spicule location. These results confirm that rotational flow field first appears mainly in the lower layers and then extends upward, indicating that the vortex structure itself has a clear upward-developing nature. The rotating spicule is gradually formed during this process. By tracking and analysing this event, we find that the spicule appears at about $t=2900$\,s and disappears at about $t=3500$\,s, corresponding to a lifetime of approximately $10$\,min. Such a numerical result is qualitatively consistent with previous analytical studies which have demonstrated that torsional motions can lead to upward plasma flows and spicule-like dynamics \citep[e.g.,][]{Oxley2020,Scalisi2021}.

The similar method as described in \citep{2022A&A...668A.118C} has been applied to identify the vortex boundaries in this work. We identify vortex structures in horizontal velocity slices. The coloured vortex boundaries shown in Figure~\ref{fig:vortex_spicule_evolution}(d)--(i) are extracted from the grid cells assigned to the selected vortex. To clearly show the vortex morphology, we only retain the main connected region in each slice, remove detached small patches, and lightly smooth the displayed vortex boundaries. The region enclosed by the vortex boundary obtained in this way is relatively small, and thus can only represent the boundary of the main central part of the vortex; some obvious rotating plasma flows in the outer region are not included. The above filtering and smoothing are used only for contour visualisation, and the original vortex identification results remain unchanged.
Figure~\ref{fig:vortex_spicule_evolution} (j) shows, at $t = 3240$\,s, an overlay of the vortex boundaries at different heights within the light blue rectangular box in the 2D slice shown in  Figure~\ref{fig:vortex_spicule_evolution}, illustrating the instantaneous three-dimensional spatial morphology of the vortex.

\begin{figure*}
\centering
\includegraphics[width=0.9\textwidth,keepaspectratio]{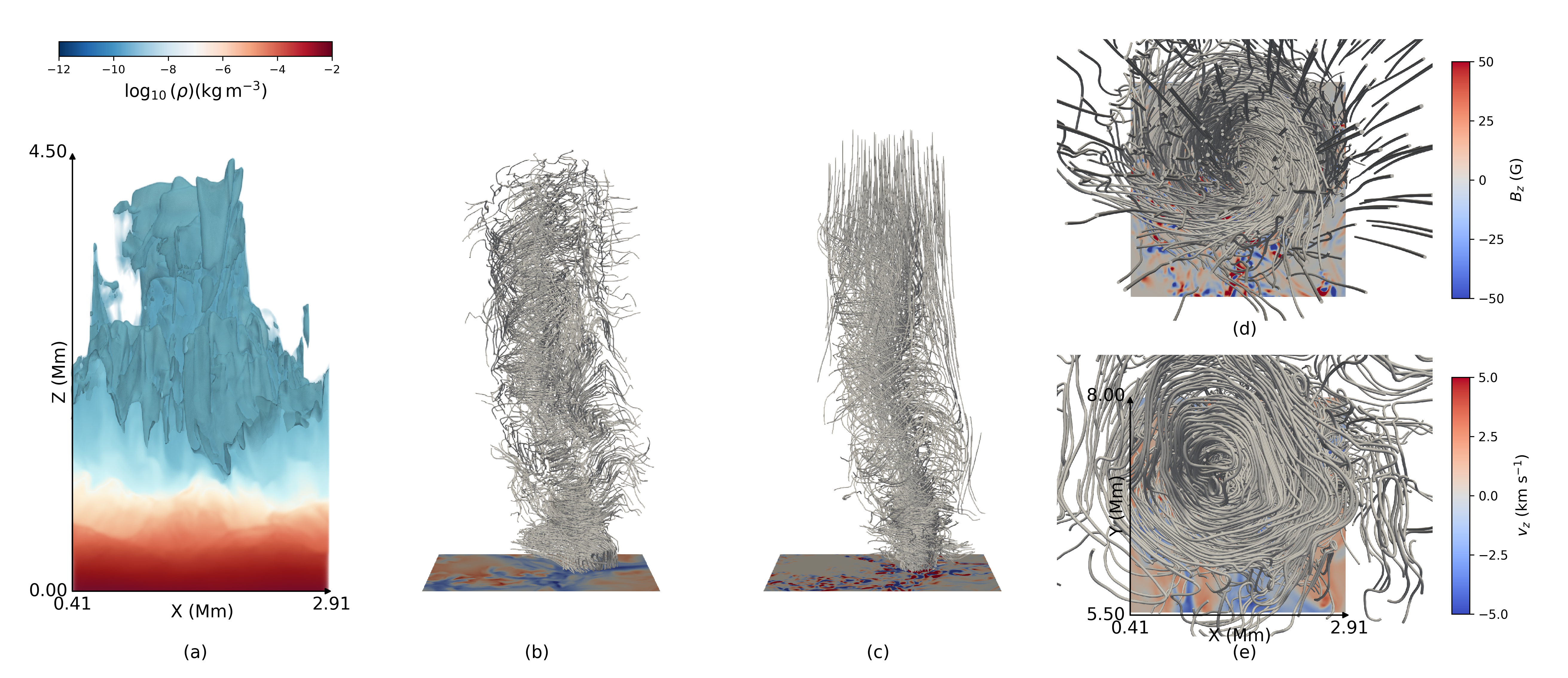}
\caption{3D views of the spicule and the magnetic tornado structure at $t=3222$\,s. 
Panel (a) shows the 3D morphology of the spicule rendered from the density distribution. 
Panels (b) and (c) show the velocity streamlines and magnetic field lines, respectively, from the same viewing angle as in panel (a). 
Panels (d) and (e) present top views of the velocity streamlines and magnetic field lines, respectively. The associated movie shows the temporal evolution of the 3D morphology of the spicule-like event, together with the velocity streamlines and magnetic field lines, from the viewing angles presented in this figure.
The associated movie is available online.}
\label{fig:3d_spicule_tornado}
\end{figure*}

Figure~\ref{fig:3d_spicule_tornado} shows the 3D structure of this event at $t=3222$\,s, highlighting the spatial relationship between the spicule and the magnetic tornado. Figure~\ref{fig:3d_spicule_tornado}(a) presents the 3D morphology of the spicule obtained from a volume rendering of the logarithmic density, with colour indicating the density distribution. The main body of the spicule is clearly seen to extend upward into the higher atmosphere. From the same view angle as Figure~\ref{fig:3d_spicule_tornado}(a), Figure~\ref{fig:3d_spicule_tornado}(b) and (c) display the 3D velocity streamlines and magnetic field lines, respectively, allowing a direct comparison between the spicule body and the surrounding flow and magnetic structures. The seed regions for these streamlines 
are selected according to the approximate position and lateral extent of the vortex at different heights. Since the vortex expands laterally with height, the seed regions are adjusted accordingly at different heights. To avoid excessive overlap and improve the clarity of the visualisation, the adopted seed radii are chosen to be slightly smaller than the visible vortex extent. Therefore, these streamlines should be regarded as representative flow lines within and around the vortex, rather than a complete filling of the entire vortex volume. To better illustrate the geometry of this rotating system. Figure~\ref{fig:3d_spicule_tornado}(d) and (e) further show top views of the velocity streamlines and magnetic field lines. These top views make it even clearer that the spicule is not an isolated vertical jet, but is embedded within a magnetic tornado, with these structures evolving together as part of the same system.

\begin{figure*}
\centering
\includegraphics[width=0.8\textwidth]{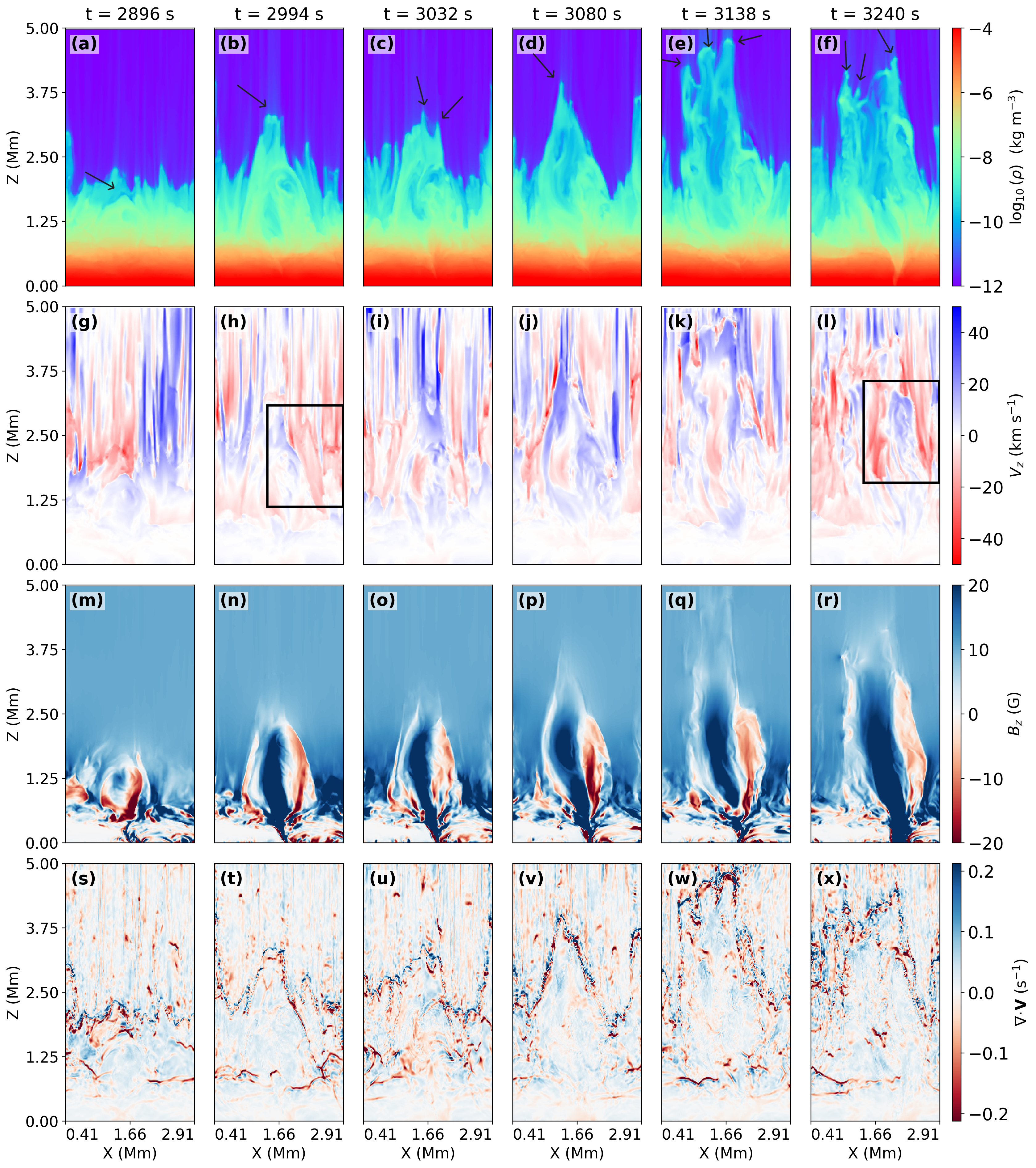}
\caption{Temporal evolution of the spicule in the $x-z$ vertical slice taken at the fixed $y$ position marked by the black dashed line in Figure~\ref{fig:vortex_spicule_evolution}(d). 
Panels (a)--(f) show the distributions of the logarithmic density ($\log_{10} (\rho)$) at the same six representative times as in Figure~\ref{fig:vortex_spicule_evolution}. 
Panels (g)--(l) show the corresponding vertical velocity slices ($V_z$). 
Panels (m)--(r) show the corresponding distributions of the vertical magnetic field ($B_z$).  Panels (s)--(x) show the corresponding distributions of the velocity divergence ($\nabla \cdot \mathbf{V}$). The two black rectangular box‌es in (h) and (l) mark the regions selected to analyse the local magnetic reconnection as shown in Figure~\ref{fig:MR_analysis}. The associated movie shows the temporal evolution of the logarithmic density, vertical velocity, vertical magnetic field, and velocity divergence.
The associated movie is available online.} 
\label{fig:xz_spicule_evolution}
\end{figure*}

Figure~\ref{fig:xz_spicule_evolution} presents the results in the $x$--$z$ vertical cross-section taken along the horizontal black dashed line shown in Figure~\ref{fig:vortex_spicule_evolution}(d). The distributions of logarithmic density ($\log_{10} (\rho)$), vertical velocity ($V_z$), vertical magnetic field ($B_z$) and divergence of velocity ($\nabla \cdot \mathbf{V}$)  are at the same six representative times as in Figure~\ref{fig:vortex_spicule_evolution}. These side-view slices reveal how the spicule-like event forms and evolves in the vertical direction. At the early stage, a pointed-top arched structure emerged at around $1.5$\,Mm (marked by the black arrow in Figure~\ref{fig:xz_spicule_evolution}(a)), while no prominent tall jet structure is present at this phase. As the event evolves, the plasma is gradually lifted upward and this event develops into a clear elongated jet-like structure. The density panels show that its morphology undergoes substantial changes during the evolution, while the corresponding velocity panels reveal complex internal motions, including both upward-moving and downward-moving plasma. 

Together, Figures~\ref{fig:vortex_spicule_evolution}--~\ref{fig:3d_spicule_tornado} and Figure~\ref{fig:xz_spicule_evolution} demonstrate from different perspectives that the event analysed in this study is a highly dynamic rotating spicule.

\subsection{Fine dynamical properties and driving mechanisms}

Figure~\ref{fig:xz_spicule_evolution}(a)--(f) shows that the spicule splits into multiple substructures, whose tops are indicated by the black arrows. Combining Figure~\ref{fig:xz_spicule_evolution}(g)--(l) with Figure~\ref{fig:vortex_spicule_evolution}(a)--(c), we find that the plasma inside the spicule exhibits highly complex motions during the overall upward displacement of the structure. Some plasma elements show upward velocities, while others move downward. In particular, the plasma in the central part of the vortex falls back at the later stage (see also in Figure~\ref{fig:vortex_spicule_evolution} (a--c) ), whereas the plasma near the vortex periphery continues to move upward at high speeds. Tracking the upward displacement of the top of the spicule-like structure gives an estimated bulk propagation speed of approximately 10--20~km\,s$^{-1}$. In contrast, Figure~\ref{fig:xz_spicule_evolution}(i--k) shows that localised upward velocities can exceed 40~km\,s$^{-1}$.

\begin{figure*}
\centering
\includegraphics[width=0.8\textwidth]{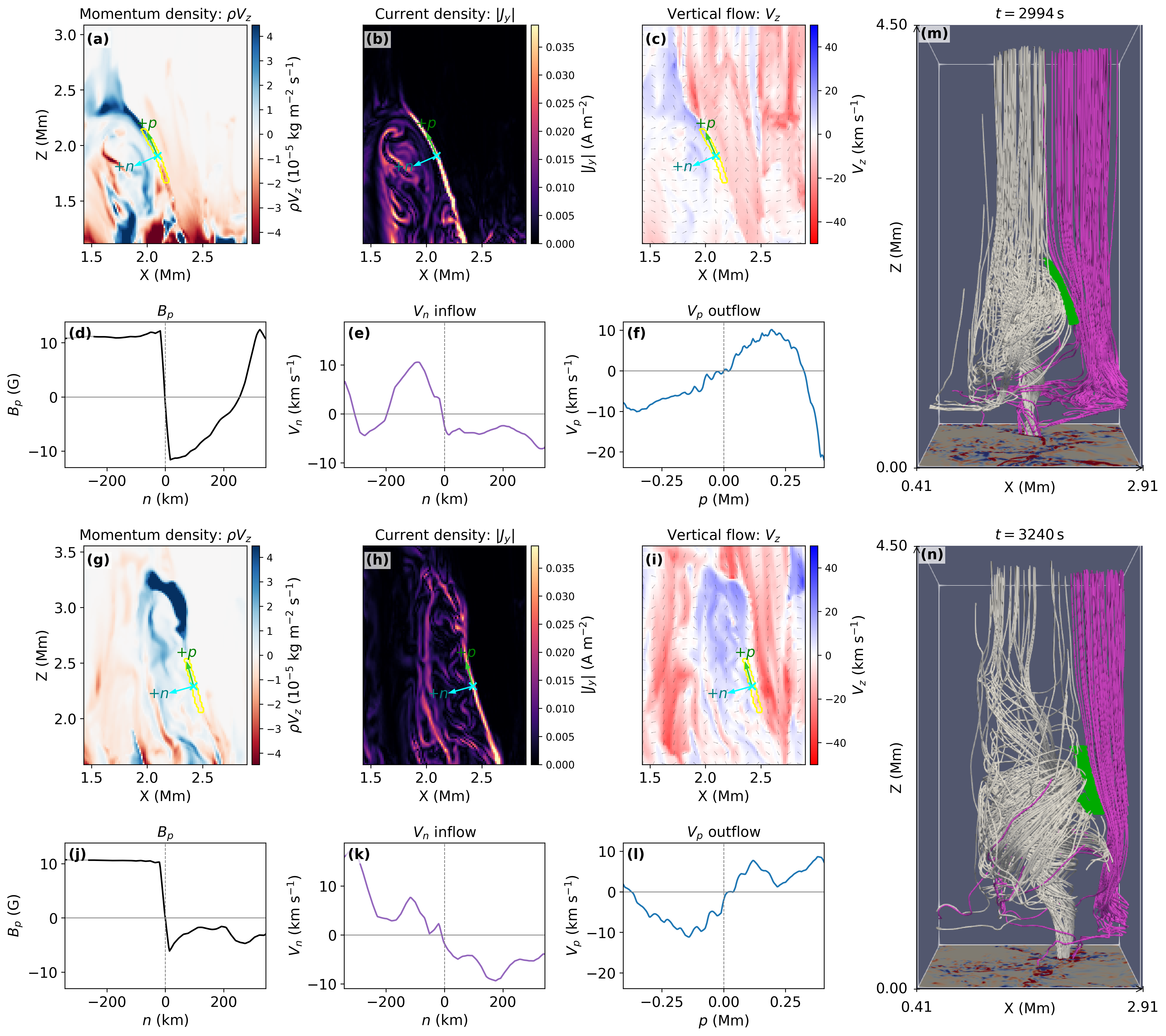}
\caption{ Panels (a--f) show the two-dimensional results of the magnetic reconnection case at $t=2994$\,s inside the black rectangular box in Figure~\ref{fig:xz_spicule_evolution} (h). Panel (m) shows the corresponding three-dimensional distributions of magnetic field lines and the isosurface of the local enhanced current density for this case at $t=2994$\,s. Panels (g--l) and panel (n) show the similar results for the other case at $t=3240$\,s inside the black rectangular box in Figure~\ref{fig:xz_spicule_evolution} (l).}
\label{fig:MR_analysis}
\end{figure*}

From the results in Figure~\ref{fig:xz_spicule_evolution}(m)--(r), one can see that the vertical magnetic field component $B_z$ is positive in the centre of the vortex, whereas negative $B_z$ (red colour) appears at the periphery of the vortex. We have carefully analysed the 3D distributions of magnetic fields in this region, and found that these negative $B_z$ components are contributed by the twisted magnetic field lines (see Figure~\ref{fig:MR_analysis} (m)--(n)) during the formation of this magnetic tornado system. The twisted magnetic field lines with negative vertical components reconnect with both external positive-polarity background fields and central positive-polarity fields. The magnetic reconnection sites rise and rotate together with the developed magnetic tornado, and many jet-like substructures in this event are closely associated with plasma upward ejections from these reconnection regions. Therefore, magnetic reconnection is also likely to be one of the main mechanisms driving the formation of this spicule-like event.

The local small scale magnetic reconnection configurations have been analysed in detail. Figure~\ref{fig:MR_analysis} shows the results in two regions marked by the black rectangular boxes in Figure~\ref{fig:xz_spicule_evolution} (h) and (l), respectively. In Figure~\ref{fig:MR_analysis} (a--c), the distributions of vertical momentum density ($\rho V_z$), absolute current density in the y-direction ($\left| J_y \right|$) and vertical velocity ($Vz$) at $t=2994$\,s in the first black rectangular box in Figure~\ref{fig:xz_spicule_evolution} (h) are presented. The thin current sheet, bi-directional outflows represent the typical characteristics of magnetic reconnection. Comparing the results in Figure~\ref{fig:MR_analysis} (a,c) and Figure~\ref{fig:xz_spicule_evolution} (b,h), one can clearly find that the upward outflow from this reconnection region contributes significantly to this spicule-like event. At the reconnection X-point, we have introduced a new coordinate system in which the p-direction is locally parallel to the thin current sheet and the n-direction is perpendicular to it. $B_p$, $V_n$ and $V_p$ represent the magnetic field along the p-direction, the velocity along the n-direction and the velocity along the p-direction, respectively. Figure~\ref{fig:MR_analysis} (d) shows the profile of $B_p$ along the n-axis through the reconnection X-point. Figure~\ref{fig:MR_analysis} (e) and (f) show the profile of $V_n$ along the n-axis and the profile of $V_p$ along the p-axis through the reconnection X-point, respectively. These profiles indicate a clear reversal of these variables across the X-point, which is also consistent with the typical characteristics of magnetic reconnection. Figure~\ref{fig:MR_analysis} (m) displays the 3D configuration of this local magnetic reconnection, the grey lines represent the magnetic field lines in the vortex system, the pink lines represent those background field lines, and the green isosurfaces between them represent the enhanced current density region. The internal magnetic reconnection between the positive and negative components of the magnetic field in the magnetic tornado system is not shown in this 3D figure. Figure~\ref{fig:MR_analysis} (g--l) and (n) in the bottom shows the similar results of the other local reconnection case at $t=3240$\,s, when is at the later stage of this spicule-like event.

In summary, the above results from Figure~\ref{fig:vortex_spicule_evolution}  to Figure~\ref{fig:MR_analysis} confirm that this spicule event is driven jointly by the vortex itself and by induced local magnetic reconnection. We also notice that clear shock structures appear both above and below this event during its formation process. In Figure~\ref{fig:xz_spicule_evolution}(s-x), the red regions denote areas with strong compression, which are corresponding to the possible shock regions. The shock above the spicule-like event is triggered by the upward flow, whereas the shock below this event is driven by the underlying convective motions and small-scale magnetic reconnection outflows. These shocks may also contribute to the formation of this spicule. This is consistent with previous two-dimensional simulation results \citep[e.g.,][]{Srivastava2025,Ni2026,Li2026}, and the detailed analysis about how the shocks driving the spicule formation has been described in those papers. 

In the following part of this subsection, we further investigate the main forces responsible for lifting the plasma upward during this event.

\begin{figure*}
\sidecaption
\includegraphics[width=12cm]{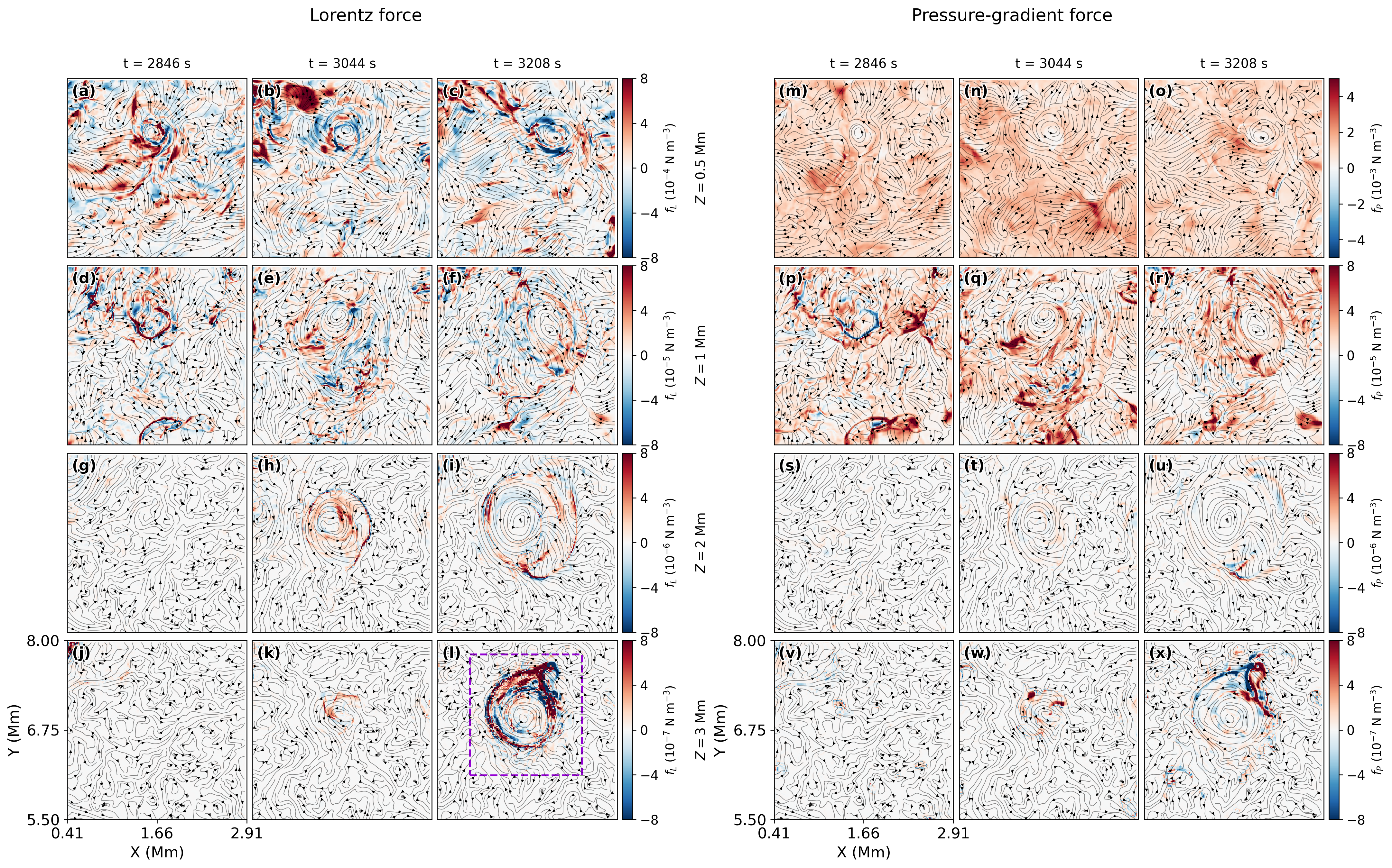}
\caption{The left panels from (a--l) show distributions of the vertical component of the Lorentz force density in the $x-y$ planes at different heights and times, $f_{\mathrm{L}}$ denotes the Lorentz force density. The right panels from (m--x) show distributions of the vertical component of the pressure-gradient force density in the $x-y$ planes at different heights and times, $f_{\mathrm{P}}$ denotes the pressure-gradient force density. The velocity streamlines are overlaid on each panel. The purple dashed rectangular box in panel (l) is the selected region to calculate the averaged values in Figure~\ref{fig:force_evolution} }
\label{fig:force_xy}
\end{figure*}

The vortex motions drive the twisting of the magnetic field lines, and a downward component of the magnetic tension force is generated in an attempt to straighten them. At the same time, as the magnetic field strength decreases with height above the solar surface, this leads to an upward magnetic pressure gradient force. The unwinding and expansion of magnetic field lines and the local reconnection processes can also cause the local enhancements of the Lorentz force. Therefore, the net vertical component of the total Lorentz force is determined by several different physical processes. Figure~\ref{fig:force_xy} (a--l) shows the distributions of the vertical component of the total Lorentz force density in horizontal $x-y$ planes at four heights, $z = 0.5$\,Mm, $1$\,Mm, $2$\,Mm, and $3$\,Mm, during the evolution process of this event. Three representative times are selected to trace how the vortex structure and its associated magnetic response are gradually established with height. Overall, the results at the lower heights of $0.5$\,Mm and $1$\,Mm are relatively complex. The regions with enhanced upward Lorentz force, shown in red, are spatially scattered, and the net upward response is not yet prominent. In contrast, at $2$\,Mm and $3$\,Mm, the upward Lorentz force gradually develops a clearer spatial organisation as the event evolves. The enhanced regions show a more evident spatial correspondence with the vortex location and are mainly distributed along the vortex periphery, where local magnetic reconnection happens (see Figure~\ref{fig:poynting_joule_bz} (e--l)). Such enhanced localised Lorentz-force structures in the middle and upper layers may also relate to the unwinding and expansion of magnetic field lines there.
 
 The results in Figure~\ref{fig:force_xy} (a--l) suggest that the magnetic response associated with the spicule becomes progressively more concentrated and structured with increasing height. This behaviour also explains the features shown in Figures~\ref{fig:xz_spicule_evolution}(g)--(l) and Figures~\ref{fig:vortex_spicule_evolution}(a)--(c): during the middle and late stages, the plasma in the central part of the vortex has already started to fall back, whereas the plasma near the vortex periphery continues to be driven upward. 

Figure~\ref{fig:force_xy} (m--x) displays the corresponding distributions of the vertical component of the pressure-gradient force density in horizontal $x-y$ planes at different heights during the same evolutionary process as in Figure~\ref{fig:force_xy} (a--l). Compared with the Lorentz force, the pressure-gradient force has a larger amplitude in the lower layers. At $z = 0.5$\,Mm, the background response is rather complex, making it difficult to directly identify a vortex-related enhanced structure. At $z = 1$\,Mm, some weak enhanced signals begin to appear around the vortex periphery, although the overall distribution remains scattered. With increasing height, the local pressure response becomes relatively clearer at $z=2$\,Mm and $3$\,Mm. However, compared with the Lorentz force, the pressure-gradient force is generally weaker in the upper layers and only shows pronounced upward enhancement at some localised positions. This result indicates that the pressure-gradient force participates in driving the formation of the spicule throughout the event, but it does not dominate at higher atmospheric layers.

\begin{figure*}
\sidecaption
\includegraphics[width=12cm]{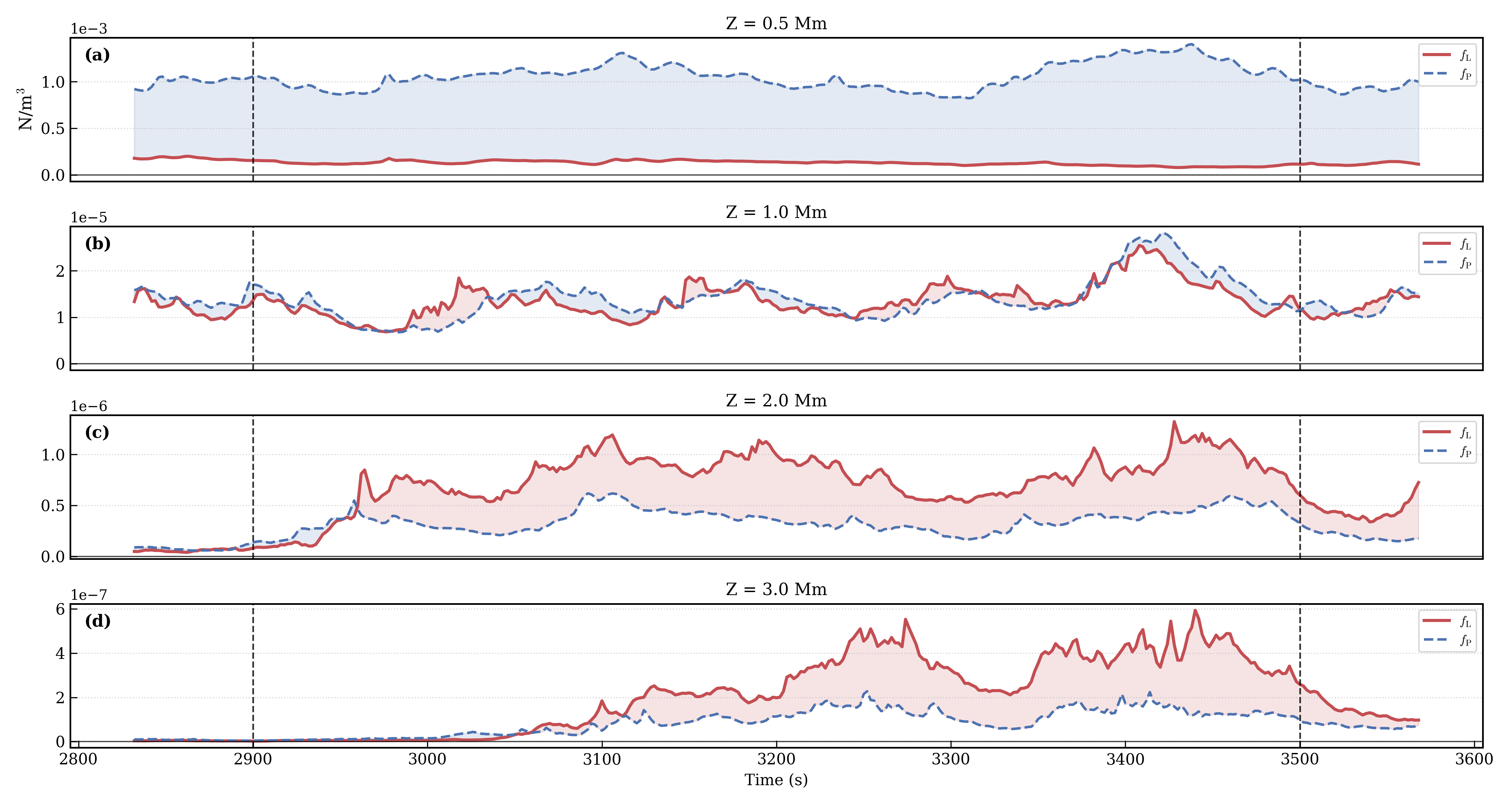}
\caption{Temporal evolution of the averaged positive vertical force-density components at different heights. 
Panels (a)--(d) correspond to the results at $z = 0.5$\,Mm, $1$\,Mm, $2$\,Mm, and $3$\,Mm, respectively. 
The red solid curves denote the Lorentz force density, and the blue dashed curves denote the pressure-gradient force density. 
The left and right vertical dashed lines approximately mark the appearance and disappearance times of the spicule, respectively. All values were obtained by averaging the positive vertical force-density components over the selected region (the purple dashed rectangular box) as shown in Figures~\ref{fig:force_xy} (l).}
\label{fig:force_evolution}
\end{figure*}

To quantitatively compare the upward forces for driving this spicule-like event at different heights, Figure~\ref{fig:force_evolution} reveals the time evolutions of the vertical components of the pressure-gradient force density and Lorentz force density at $z = 0.5$\,Mm, $1$\,Mm, $2$\,Mm, and $3$\,Mm. For each time step, the values in the curves are obtained by averaging the positive vertical force-density components over the selected region, $0.859$\,Mm\,$<x<2.422$\,Mm and $6.125$\,Mm\,$<y<7.813$\,Mm (in the purple dashed rectangular box as shown in Figures~\ref{fig:force_xy} (l)).

The curves reveal clear height-dependent differences in the force response. At the lower height of $z = 0.5$\,Mm, the upward pressure-gradient force plays a clear dominant role, whereas the upward Lorentz force remains relatively weak. At $z = 1$\,Mm, a more pronounced magnetic response begins to appear, but the pressure gradient force still makes a significant contribution. As the height increases to $2$\,Mm and $3$\,Mm, the upward Lorentz force becomes significantly larger than the upward pressure gradient force, and exhibits a more pronounced trend of continuous enhancement during the event development phase. This result is consistent with those shown in Figures~\ref{fig:force_xy}.  Such height-dependent stratification of the driving forces is a key feature for understanding the formation of this event.

\subsection{Mass and energy transport}

\begin{figure*}
\sidecaption
\includegraphics[width=12cm]{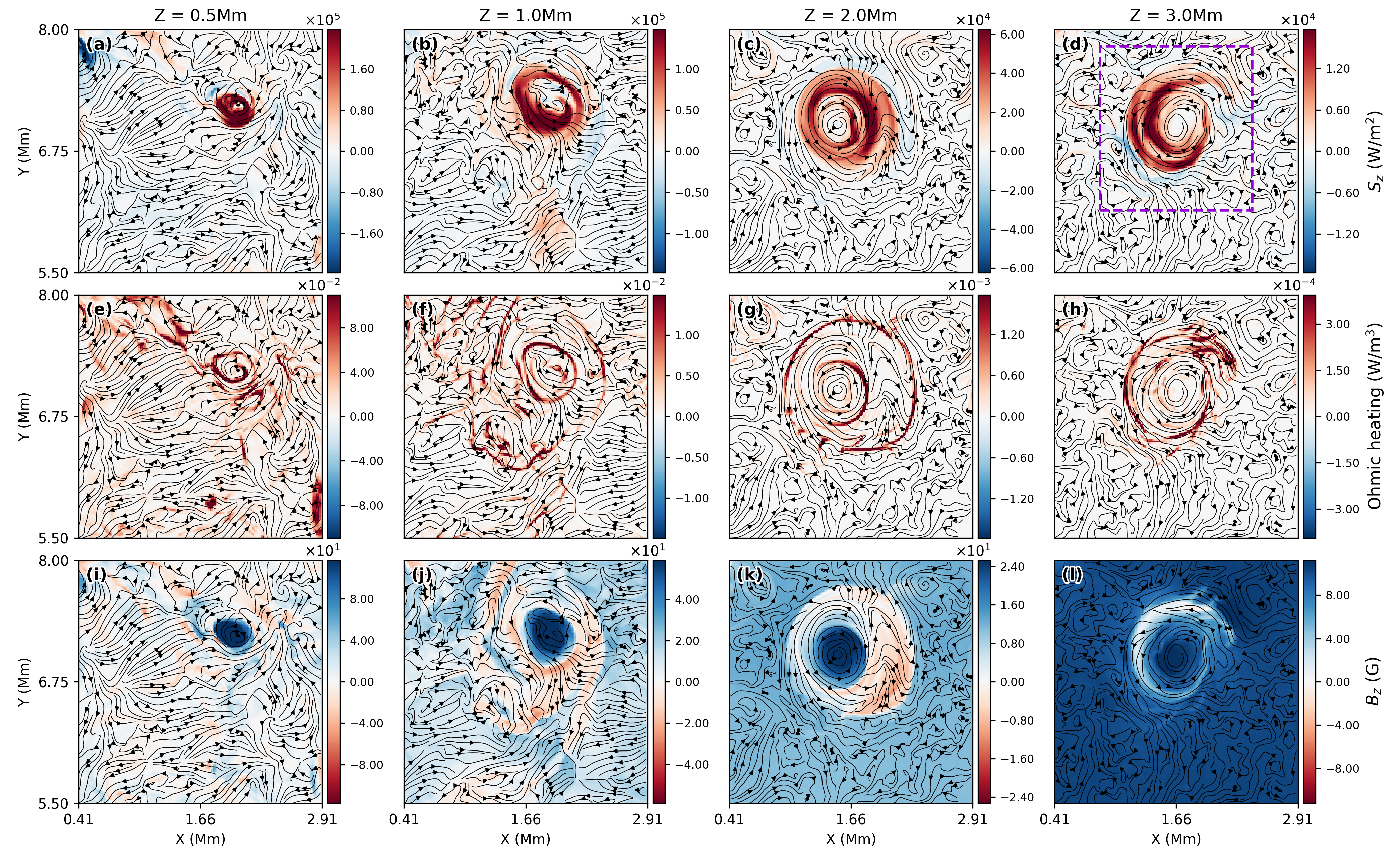}
\caption{Distributions of the vertical Poynting flux density, Joule heating rate, and vertical magnetic field at $t=3188$~s. 
Panels (a)--(d) show the vertical Poynting flux density at heights of $0.5$\,Mm, $1$\,Mm, $2$\,Mm, and $3$\,Mm, respectively. Panels (e)--(h) show the corresponding heating rate per unit volume due to Joule dissipation. Panels (i)--(l) show the corresponding distributions of the vertical magnetic field $B_z$. 
The black curves with arrows indicate the magnetic field lines. The purple dashed rectangular box in panel (d) is the selected region to calculate the averaged Poynting flux density and mass flux density in Figure~\ref{fig:poynting_mass_evolution}}
\label{fig:poynting_joule_bz}
\end{figure*}

In this section, we focus on the transport of mass and energy between different atmospheric layers during the formation and evolution of this event. Figure~\ref{fig:poynting_joule_bz} shows the distributions of the vertical Poynting flux density, Joule heating rate per unit volume, and vertical magnetic field strength in the $x-y$ planes at four different heights during the intermediate stage of the spicule formation.

The first row of Figure~\ref{fig:poynting_joule_bz} shows that the vertical Poynting flux is not uniformly distributed over the analysed region, but is strongly enhanced in localised regions associated with the vortex. From the photosphere to the upper chromosphere, these regions exhibit substantial upward Poynting flux. At $z=3$\,Mm, the vertical Poynting flux density can reach values of the order of $10^{4}~{\rm W~m^{-2}}$. By comparing the second and third rows of Figure~\ref{fig:poynting_joule_bz}, we find that Joule heating mainly occurs in regions where positive and negative polarity vertical magnetic fields approach each other. These heating regions are concentrated at the local magnetic reconnection regions, where the twisted magnetic field lines with negative vertical components separately reconnect with external positive-polarity background fields and central positive-polarity fields. As shown in Figure~\ref{fig:poynting_joule_bz}(g), these regions appear as two nested ring-like structures of different sizes. Such results indicate that magnetic reconnection significantly heats the local plasma in this process.

\begin{figure*}
\sidecaption
\includegraphics[width=12cm]{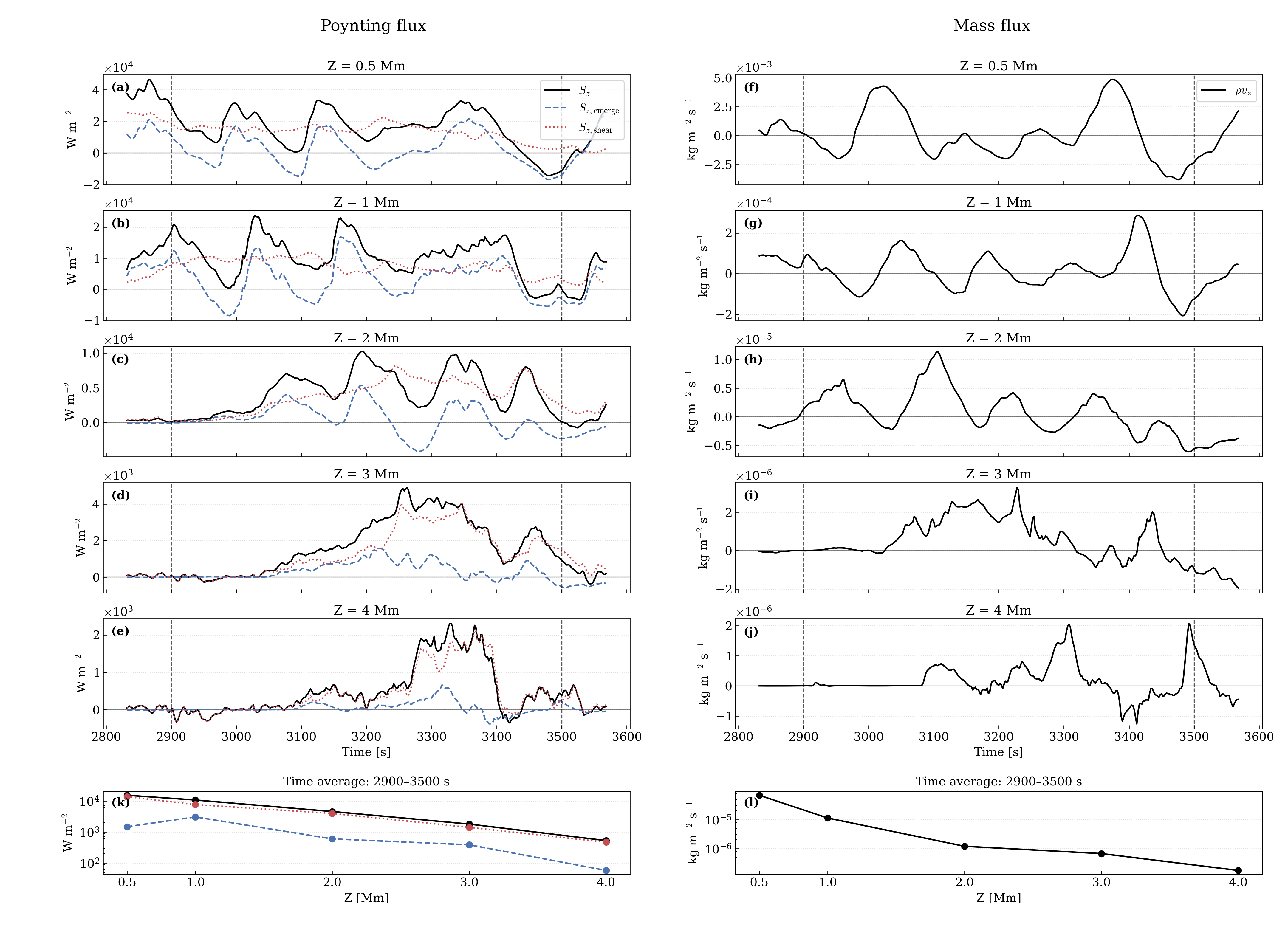}
\caption{Panels (a)--(e) correspond to the averaged vertical Poynting flux densities at $z = 0.5, 1, 2, 3$, and $4$~Mm, and all the curves at each time point are calculated in the selected region (the purple dashed rectangular box as shown in panel (d) of Figure~\ref{fig:poynting_joule_bz}). The black solid curves denote the total vertical Poynting flux density, while the blue dashed and red dotted curves represent the emergence and shear terms, respectively. Panel (k) shows the averaged vertical Poynting flux densities at five different heights during the period from $2900$\,s to $3500$\,s, and the black, blue and red dots respectively correspond to the total Poynting flux density, the emergence and shear terms. Panels (f)--(j) correspond to the vertical mass flux densities at $z = 0.5$\,Mm, $1$\,Mm, $2$\,Mm, $3$\,Mm, and $4$\,Mm, respectively, and all curves are averaged over the same selected region as calculating the averaged vertical Poynting flux density. Panel (l) shows the averaged vertical mass flux densities at five different heights during the period from $2900$\,s to $3500$\,s.} 
\label{fig:poynting_mass_evolution}
\end{figure*}

Following previous studies of magnetic-energy transport through the photosphere, the vertical Poynting flux density can be decomposed into a shear term and an emergence term \citep[e.g.,][]{Magara2003,Welsch2015} as

\begin{equation}
S_z = S_{z,{\rm shear}} + S_{z,{\rm emerge}},
\end{equation}
where
\begin{equation}
S_{z,{\rm shear}} = -\frac{1}{\mu_0} B_z \left( v_x B_x + v_y B_y \right),
\end{equation}
and
\begin{equation}
S_{z,{\rm emerge}} = \frac{1}{\mu_0} v_z \left( B_x^2 + B_y^2 \right).
\end{equation}
The shear term ($S_{z,{\rm shear}}$) is associated with the horizontal shearing or twisting motion of the vertical magnetic field, whereas the emergence term ($S_{z,{\rm emerge}}$) is related to the vertical transport of horizontal magnetic field. These two terms are commonly interpreted as the energy transport caused by shear motions and magnetic flux emergence, respectively.

To quantitatively characterise the energy transport in the rotating spicule-like event, Figure~\ref{fig:poynting_mass_evolution} (a--e) plots the evolution of the averaged vertical Poynting flux densities at five heights, $z = 0.5$\,Mm, $1$\,Mm, $2$\,Mm, $3$\,Mm, and $4$\,Mm, over the region $0.859$\,Mm\,$<x<2.422$\,Mm and $6.125$\,Mm\,$<y<7.813$\,Mm (in the purple dashed rectangular box as shown in panel (d) of Figure~\ref{fig:poynting_joule_bz}). The different components of the Poynting flux density are also presented. These plots are used to compare the strength of electromagnetic energy transport at different heights and its temporal evolution. The curves indicate that, during the formation process of this event, the positive vertical Poynting flux density is clearly enhanced at all five heights. Although the magnitude and detailed temporal behaviour of the enhancements differ from height to height, the overall trend is consistent: as the vortex develops and the spicule forms, the upward Poynting flux is enhanced in all atmospheric layers. Such a result is consistent with those shown in Figure~\ref{fig:poynting_joule_bz}. From Figure~\ref{fig:poynting_mass_evolution}(k), one can find that during the period when the spicule exists, from $t=2900$\,s to $t=3500$\,s, the mean Poynting flux density at a height of $z = 4$\,Mm is approximately $525$\,W\,m$^{-2}$. These plots also indicate that the shearing term  $S_{z,{\rm shear}}$ (the red dots) plays the dominant role in transporting electromagnetic energy upward at all altitudes. The results confirm that this event transports a considerable amount of electromagnetic energy into the low corona and that the twisted motions caused by the vortex play a leading role in this process.

We also investigate the mass transport caused by this spicule event. Figure~\ref{fig:poynting_mass_evolution} (f--j) depicts the time evolutions of the averaged vertical mass flux density at the five heights over the same selected region as calculating the averaged vertical Poynting flux density. The average mass flux density varies with time at different heights, alternating between positive and negative values, while its magnitude decreases as the height increases. By further averaging over the interval from $t=2900$\,s to $t=3500$\,s, as shown in Figure~\ref{fig:poynting_mass_evolution} (l), one can find that the mean mass flux densities are positive at all the altitudes. The value reaches about $10^{-7}$\, kg\,m$^{-2}$\,s$^{-1}$ at a height of $4$\,Mm above the solar surface. Such results suggest that a considerable amount of plasma from the lower atmosphere is transported into the low corona during the formation process of this spicule-like event.

During the formation process of this event, the plasma and magnetic field are continuously disturbed, leading to the generation of different types of MHD waves. Following the methods of \citet{Vigeesh2012}, \citet{Khomenko2012}, and \citet{Mumford2015}, we perform the wave-mode decomposition method to estimate the average energy flux density carried by different wave modes. The estimated energy flux densities are given by
\begin{equation}
F_{\parallel} = \rho \  v_{\parallel}^{2} c_{\rm s},
\end{equation}
\begin{equation}
F_{\perp} = \rho v_{\perp}^{2} V_{\rm A},
\end{equation}
and
\begin{equation}
F_{\phi} = \rho v_{\phi}^{2} V_{\rm A}.
\end{equation}
Here, $F_{\parallel}$, $F_{\perp}$, and $F_{\phi}$ denote the energy flux densities parallel to the magnetic field, perpendicular to the magnetic field, and in the azimuthal direction, respectively. The corresponding velocity perturbations are $v_{\parallel}$, $v_{\perp}$, and $v_{\phi}$. The quantities $c_{\rm s}$ and $V_{\rm A}$ are the sound speed and Alfv\'en speed, respectively. In the low-$\beta$ regime, where $\beta \ll 1$, $F_{\parallel}$, $F_{\perp}$, and $F_{\phi}$ can be regarded as estimates of the energy flux densities carried by the slow, fast, and Alfv\'en modes, respectively. This method defines an available wave energy flux density as the product of the perturbation energy density and the corresponding characteristic wave speed. It therefore provides a more stable measure of the average upward energy transport carried by waves. Compared with directly averaging the oscillatory wave energy flux in time, this treatment avoids cancellation between positive and negative contributions caused by standing waves or phase variations, and is therefore more suitable for comparing the overall efficiency of energy transport by different wave modes.

\begin{figure*}
\sidecaption
\includegraphics[width=12cm]{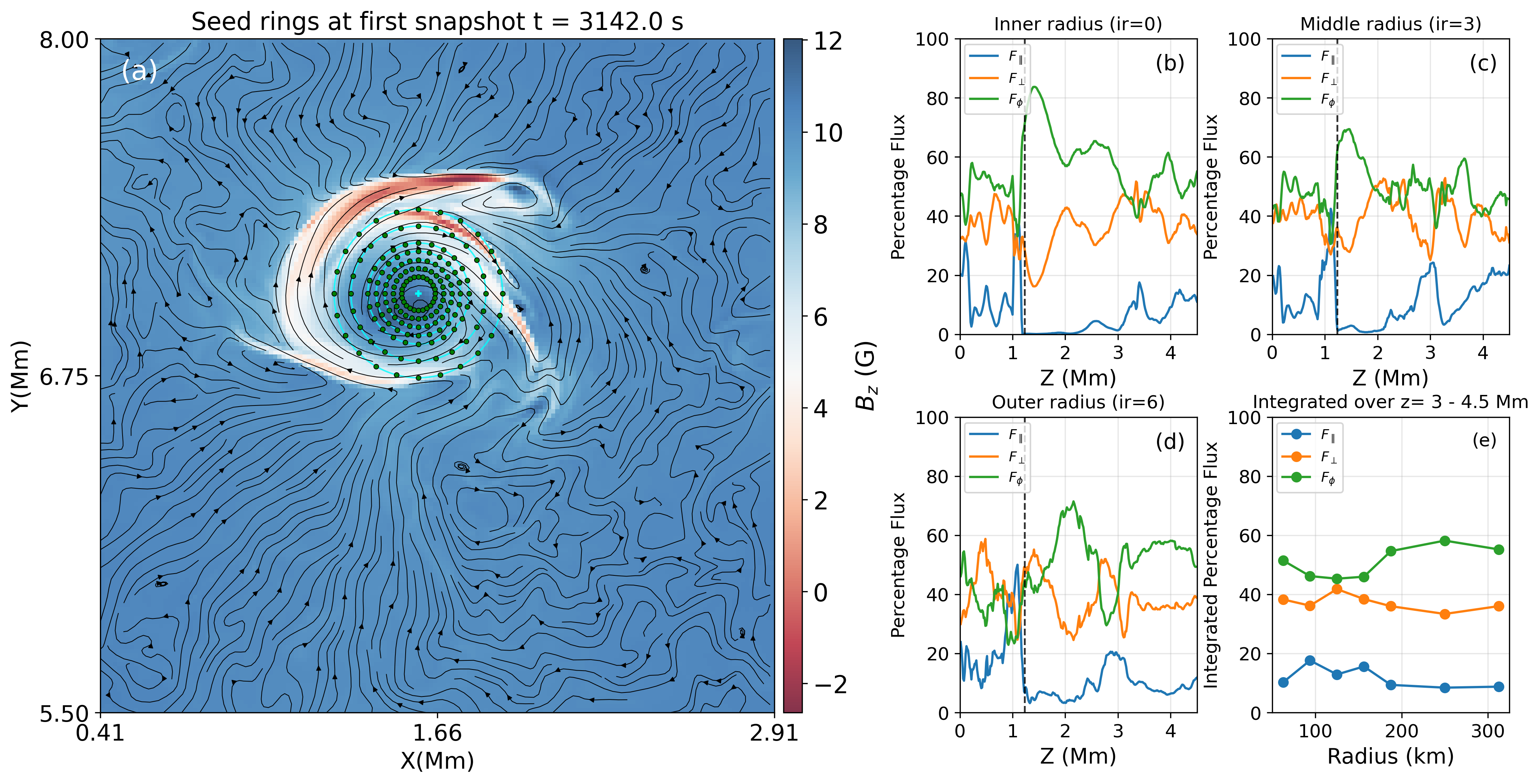}
\caption{Wave-mode analysis results based on magnetic flux surfaces. 
Panel (a) shows the seven seed circles with different radii used to construct the flux surfaces, overplotted on the vertical magnetic field and magnetic field lines at a height of 3~Mm. 
Panels (b)--(d) show the variation with height of the proportions of energy flux density in different directions for the seed circles with radius indices $r=0$, $r=3$, and $r=6$, respectively. 
The green, blue, and orange curves represent the energy flux densities parallel to the magnetic field, perpendicular to the magnetic field, and in the azimuthal direction, respectively. 
Panel (e) shows the three proportions at different radii, obtained by integrating over the height range $3~{\rm Mm}<z<4.5~{\rm Mm}$. 
The vertical dashed lines in panels (b)--(d) indicate the heights where the mean plasma $\beta$ is equal to unity.}
\label{fig:wave_mode_decomposition}
\end{figure*}

Figure~\ref{fig:wave_mode_decomposition} presents the wave-mode analysis averaged values which are calculated from $t=3142$\,s to $t=3152$\,s. All quantities in this figure show the temporal averages over this short interval. The left panel marks the seven seed circles with different radii used to construct the magnetic flux surfaces. Figures~\ref{fig:wave_mode_decomposition}(b)--(d) respectively show the variations with height of the proportions of energy flux densities in different directions within the inner, middle, and outer rings. The vertical dashed lines mark the locations where the mean plasma $\beta$ is equal to unity. Figure~\ref{fig:wave_mode_decomposition}(e), presents the three proportions at different radii. These values are obtained by integrating over the height range of $3~{\rm Mm} < z < 4.5~{\rm Mm}$, where the plasma $\beta$ is much smaller than unity. The results indicate that, during the representative stage analysed here, the energy transport is dominated by the Alfv\'en mode at all radii, from the inner to the outer seed circles. Based on the results in Figure~\ref{fig:wave_mode_decomposition}(e), we can find that the Alfv\'en waves, fast-mode waves, and slow-mode waves carry energy into the lower corona in an approximate ratio of $10:7:3$.

The vortex structure undergoes significant drift and morphological changes over the course of its evolution. Therefore, we only choose a short time interval of ten seconds to calculate these energy flux densities based on the fixed seed circles. The vortex structure is at this maturity stage, and its morphology and position change very little during this short period, which guarantees the accuracy for calculating those results presented in Figure~\ref{fig:wave_mode_decomposition}. Sampling was performed separately on rings at several different radii, with the same number of magnetic field lines selected on each ring. The average at each radius was first calculated, and the results for the different radii were then averaged directly. Therefore, the inner and outer rings have the same weight in the final statistics. Thus, the fact that the areas at different radii are not equal has no effect on the results. The selected radii were intentionally restricted to the coherent vortex region, and therefore the outermost radius is still located inside the rotating magnetic structure rather than representing the ambient background.

\section{Conclusions and discussion}

In this paper, a 3D RMHD simulation was performed to investigate the relationship between vortex and spicule-like event in the lower solar atmosphere. We focused on one representative event, and analysed its driving mechanism, morphological evolution, 3D spatial correspondence, dynamical response, upward energy and mass transport, and wave properties. The main conclusions are summarised as follows.

\begin{enumerate}
\item The rotating spicule-like event is driven jointly by a vortex and the induced multiple small scale magnetic reconnection outflows. Prior to the formation of this event, a photospheric vortex appears and twists the magnetic field lines, then develops upward into higher atmospheric layers, forming a magnetic tornado system. The rotating spicule-like event is generated during this process. At the periphery of this magnetic tornado, twisted magnetic field lines with negative vertical components reconnect with both external positive-polarity background fields and central positive-polarity fields. The reconnection sites rise and rotate together with the developed magnetic tornado, and many jet-like substructures in this event are closely associated with plasma ejections from these reconnection regions.

\item Analysis of the vertical force components on horizontal slices at different heights shows that the main force responsible for lifting the spicular plasma is strongly height dependent. In the photosphere and lower chromosphere, the pressure-gradient force dominates the upward driving, whereas the Lorentz force gradually becomes the dominant upward force in the upper chromosphere. The enhanced Lorentz force is mainly concentrated around the periphery of the vortex, where magnetic reconnection often occurs and magnetic field lines may also unwind and expand.

\item In the region where the vortex and spicule-like event develop, the upward vertical Poynting flux and mass flux are significantly enhanced. The shearing and twisting motions of the magnetic field play the leading role in transporting electromagnetic energy upward at all altitudes. In the low corona at a height of $4$\,Mm above the solar surface, the mean vertical Poynting flux density during the spicule formation is approximately $525$\,W\,m$^{-2}$, and the mean mass flux density is about $10^{-7}$\,kg\,m$^{-2}$\,s$^{-1}$.

\item A wave-mode decomposition of the MHD waves generated during this rotating spicule-like event shows that the wave energy transport is dominated overall by the Alfv\'en mode. Alfv\'en waves, fast-mode waves, and slow-mode waves carry energy into the lower corona in an approximate ratio of $10:7:3$.
\end{enumerate}

Based on observational and numerical studies, a variety of physical mechanisms have been proposed to account for the generation of solar spicules. The physical process of vortex-driven spicule formation has been successfully modelled/captured in previous 3D MHD simulations \citep{Iijima2017,Dey2026}. Our simulation results make a step forward as we show that a spicule-like event can be driven by the combined action of vortex motions and the induced multiple small scale reconnection outflows. The recent 2.5D MHD simulations have proved that shocks and magnetic reconnection can jointly drive the spicule formation \citep[e.g.,][]{Ni2026,Li2026}. In addition, recent observational and numerical studies have suggested that torsional motions and shocks can also act together to trigger the formation of spicules \citep[e.g.,][]{Dey2022,Lee2026}. Therefore, the physical processes responsible for spicule formation, as well as the fine structure of spicules, are much more complex than previously thought and often involve the combined action of multiple driving mechanisms.

We should also point out that the reconnection configuration for this particular event is not the same as the previous works about magnetic reconnection between the emerged twisted flux rope and the background magnetic field \citep[e.g.,][]{2013ApJ...771...20M}, that triggers the formation of a solar jet. The background magnetic field in the corona is indeed unipolar both in our work and \cite{2013ApJ...771...20M}. However, the twisted magnetic flux ropes that participate in magnetic reconnection in these two simulations are very different. In \cite{2013ApJ...771...20M}, the emerged twisted magnetic flux rope is initially oriented along the y-axis, that is parallel to the solar surface. The rotating coronal jet caused by reconnection is generated not along but above the original twisted flux rope, with a completely different axis. In contrast, the twisted magnetic flux rope in our simulation is caused by the vortex motion in the lower atmosphere and it is oriented along the z-axis, that is perpendicular to the solar surface. The rotating spicule-like event forms gradually within the magnetic tornado and shares its z-direction rotation axis. 

Our results further demonstrate that, during the spicule formation, a considerable amount of electromagnetic energy and plasma can be transported into the upper atmosphere. Such energy and mass fluxes may provide important sources for maintaining the high temperature of the corona and supplying material to the solar wind. In this work, the mean mass flux density measured at the same altitude in the low corona is more than one order of magnitude higher than the values reported in previous studies \citep{Ni2026,Chen2025}. This is because we measured this value during a single spicule formation process, whereas those previous studies calculated the mean value over the entire simulation domain, which included different jets or spicules and regions without such events. In the future, a high-resolution 3D numerical simulation on a much larger scale, extending to higher coronal heights will be required to evaluate the contribution of spicules to coronal heating and the origin of the solar wind under realistic solar conditions.

This work provides the first assessment of the ratios at which Alfv\'en waves, fast-mode waves, and slow-mode waves, excited during a rotating spicule formation process, carry energy into the lower corona. It should be noted, however, that the vortex structure undergoes significant drift and morphological changes during its evolution. Therefore, the current analysis based on fixed seed circles can only provide relatively accurate results over a short time interval, and cannot fully trace the complete 3D evolution of the entire event. Future studies should combine longer time series, automatic structure tracking, and statistical analyses of multiple events.

We emphasize that the present study focuses primarily on the dynamical evolution of a spicule-like event and its driving mechanisms, rather than on establishing a direct observational correspondence. A direct comparison with the observed spicules through synthetic diagnostics in different spectral lines and passbands will be pursued in future work.

\begin{acknowledgements}
We would like to thank Robert Cameron and Damien Przybylski for their advice with the MURaM code, and Sami K. Solanki for his helpful discussions, which improved this work. This research is supported by the Strategic Priority Research Program of the Chinese Academy of Sciences under grant No. XDB0560000; the NSFC grants 12373060; the National Key R\&D Program of China grant Nos. 2022YFF0503804 ( 2022YFF0503800) and 2022YFF0503003 ( 2022YFF0503000) ; the Basic Research of Yunnan Province in China under grant No. 202401AS070044; China’s Space Origins Exploration Program; Yunnan Key Laboratory of Solar Physics and Space Science under grant No. 202205AG070009;  This work was also supported by the International Space Science Institute project ( ISSI-BJ ID 24-604)  on “Small-scale eruptions in the Sun.” R.E. acknowledges the NKFIH (OTKA, grant No.K142987) Hungary for enabling this research. R.E. is grateful to Science and Technology Facilities Council (STFC, grant No. ST/ M000826/ 1) UK, PIFI ( China, grant No. 2024PVA0043) , and the NKFIH Excellence grant TKP2021-NKTA-64 (Hungary). The simulation work was carried out at the National Supercomputer Center in Tianjin, and the calculations were performed on the Tianhe new-generation supercomputer. The numerical data analysis was performed at the Computational Solar Physics Laboratory of Yunnan Observatories.
\end{acknowledgements}

\bibliographystyle{aa}
\bibliography{refs}

@ARTICLE{Brandt1988,
       author = {{Brandt}, P.~N. and {Scharmer}, G.~B. and {Ferguson}, S. and {Shine}, R.~A. and {Tarbell}, T.~D. and {Title}, A.~M.},
        title = "{Vortex flow in the solar photosphere}",
      journal = {\nat},
         year = 1988,
        month = sep,
       volume = {335},
       number = {6187},
        pages = {238-240},
          doi = {10.1038/335238a0},
       adsurl = {https://ui.adsabs.harvard.edu/abs/1988Natur.335..238B}
}

@ARTICLE{Bonet2008,
       author = {{Bonet}, J.~A. and {M{\'a}rquez}, I. and {S{\'a}nchez Almeida}, J. and {Cabello}, I. and {Domingo}, V.},
        title = "{Convectively Driven Vortex Flows in the Sun}",
      journal = {\apjl},
         year = 2008,
        month = nov,
       volume = {687},
       number = {2},
        pages = {L131},
          doi = {10.1086/593329},
archivePrefix = {arXiv},
       eprint = {0809.3885},
 primaryClass = {astro-ph},
       adsurl = {https://ui.adsabs.harvard.edu/abs/2008ApJ...687L.131B}
}

@ARTICLE{Xie2025,
       author = {{Xie}, Quan and {Liu}, Jiajia and {Nelson}, Chris J. and {Erd{\'e}lyi}, Robert and {Wang}, Yuming},
        title = "{Photospheric Swirls in a Quiet-Sun Region}",
      journal = {\apj},
         year = 2025,
        month = jan,
       volume = {979},
       number = {1},
          eid = {27},
        pages = {27},
          doi = {10.3847/1538-4357/ad99d4},
archivePrefix = {arXiv},
       eprint = {2412.03816},
 primaryClass = {astro-ph.SR},
       adsurl = {https://ui.adsabs.harvard.edu/abs/2025ApJ...979...27X}
}

@ARTICLE{Vargas2011,
       author = {{Vargas Dom{\'\i}nguez}, S. and {Palacios}, J. and {Balmaceda}, L. and {Cabello}, I. and {Domingo}, V.},
        title = "{Spatial distribution and statistical properties of small-scale convective vortex-like motions in a quiet-Sun region}",
      journal = {\mnras},
         year = 2011,
        month = sep,
       volume = {416},
       number = {1},
        pages = {148-154},
          doi = {10.1111/j.1365-2966.2011.19048.x},
archivePrefix = {arXiv},
       eprint = {1105.3092},
 primaryClass = {astro-ph.SR},
       adsurl = {https://ui.adsabs.harvard.edu/abs/2011MNRAS.416..148V}
}

@ARTICLE{Wedemeyer2009,
       author = {{Wedemeyer-B{\"o}hm}, S. and {Rouppe van der Voort}, L.},
        title = "{Small-scale swirl events in the quiet Sun chromosphere}",
      journal = {\aap},
         year = 2009,
        month = nov,
       volume = {507},
       number = {1},
        pages = {L9-L12},
          doi = {10.1051/0004-6361/200913380},
archivePrefix = {arXiv},
       eprint = {0910.2226},
 primaryClass = {astro-ph.SR},
       adsurl = {https://ui.adsabs.harvard.edu/abs/2009A&A...507L...9W}
}

@ARTICLE{Park2016,
       author = {{Park}, S.-H. and {Tsiropoula}, G. and {Kontogiannis}, I. and {Tziotziou}, K. and {Scullion}, E. and {Doyle}, J.~G.},
        title = "{First simultaneous SST/CRISP and IRIS observations of a small-scale quiet Sun vortex}",
      journal = {\aap},
         year = 2016,
        month = feb,
       volume = {586},
          eid = {A25},
        pages = {A25},
          doi = {10.1051/0004-6361/201527440},
archivePrefix = {arXiv},
       eprint = {1512.06032},
 primaryClass = {astro-ph.SR},
       adsurl = {https://ui.adsabs.harvard.edu/abs/2016A&A...586A..25P}
}

@ARTICLE{Shetye2019,
       author = {{Shetye}, Juie and {Verwichte}, Erwin and {Stangalini}, Marco and {Judge}, Philip G. and {Doyle}, J.~G. and {Arber}, Tony and {Scullion}, Eamon and {Wedemeyer}, Sven},
        title = "{Multiwavelength High-resolution Observations of Chromospheric Swirls in the Quiet Sun}",
      journal = {\apj},
         year = 2019,
        month = aug,
       volume = {881},
       number = {1},
          eid = {83},
        pages = {83},
          doi = {10.3847/1538-4357/ab2bf9},
       adsurl = {https://ui.adsabs.harvard.edu/abs/2019ApJ...881...83S}
}

@ARTICLE{Dakanalis2022,
       author = {{Dakanalis}, I. and {Tsiropoula}, G. and {Tziotziou}, K. and {Kontogiannis}, I.},
        title = "{Chromospheric swirls. I. Automated detection in H{\ensuremath{\alpha}} observations and their statistical properties}",
      journal = {\aap},
         year = 2022,
        month = jul,
       volume = {663},
          eid = {A94},
        pages = {A94},
          doi = {10.1051/0004-6361/202243236},
archivePrefix = {arXiv},
       eprint = {2205.07720},
 primaryClass = {astro-ph.SR},
       adsurl = {https://ui.adsabs.harvard.edu/abs/2022A&A...663A..94D}
}

@ARTICLE{Pike1998,
       author = {{Pike}, C.~D. and {Mason}, H.~E.},
        title = "{Rotating Transition Region Features Observed with the SOHO Coronal Diagnostic Spectrometer}",
      journal = {\solphys},
         year = 1998,
        month = oct,
       volume = {182},
       number = {2},
        pages = {333-348},
          doi = {10.1023/A:1005065704108},
       adsurl = {https://ui.adsabs.harvard.edu/abs/1998SoPh..182..333P}
}

@ARTICLE{Li2012,
       author = {{Li}, Xing and {Morgan}, Huw and {Leonard}, Drew and {Jeska}, Lauren},
        title = "{A Solar Tornado Observed by AIA/SDO: Rotational Flow and Evolution of Magnetic Helicity in a Prominence and Cavity}",
      journal = {\apjl},
         year = 2012,
        month = jun,
       volume = {752},
       number = {2},
          eid = {L22},
        pages = {L22},
          doi = {10.1088/2041-8205/752/2/L22},
archivePrefix = {arXiv},
       eprint = {1205.3819},
 primaryClass = {astro-ph.SR},
       adsurl = {https://ui.adsabs.harvard.edu/abs/2012ApJ...752L..22L}
}

@ARTICLE{Liu2012,
       author = {{Liu}, Jiajia and {Zhou}, Zhenjun and {Wang}, Yuming and {Liu}, Rui and {Wang}, Bin and {Liao}, Chijian and {Shen}, Chenglong and {Zheng}, Huinan and {Miao}, Bin and {Su}, Zhenpeng and {Wang}, S.},
        title = "{Slow Magnetoacoustic Waves Observed above a Quiet-Sun Region in a Dark Cavity}",
      journal = {\apjl},
         year = 2012,
        month = oct,
       volume = {758},
       number = {2},
          eid = {L26},
        pages = {L26},
          doi = {10.1088/2041-8205/758/2/L26},
archivePrefix = {arXiv},
       eprint = {1209.3370},
 primaryClass = {astro-ph.SR},
       adsurl = {https://ui.adsabs.harvard.edu/abs/2012ApJ...758L..26L}
}

@ARTICLE{Su2012,
       author = {{Su}, Yang and {Wang}, Tongjiang and {Veronig}, Astrid and {Temmer}, Manuela and {Gan}, Weiqun},
        title = "{Solar Magnetized ``Tornadoes:'' Relation to Filaments}",
      journal = {\apjl},
         year = 2012,
        month = sep,
       volume = {756},
       number = {2},
          eid = {L41},
        pages = {L41},
          doi = {10.1088/2041-8205/756/2/L41},
archivePrefix = {arXiv},
       eprint = {1208.0138},
 primaryClass = {astro-ph.SR},
       adsurl = {https://ui.adsabs.harvard.edu/abs/2012ApJ...756L..41S}
}

@ARTICLE{Wedemeyer2012,
       author = {{Wedemeyer-B{\"o}hm}, Sven and {Scullion}, Eamon and {Steiner}, Oskar and {Rouppe van der Voort}, Luc and {de La Cruz Rodriguez}, Jaime and {Fedun}, Viktor and {Erd{\'e}lyi}, Robert},
        title = "{Magnetic tornadoes as energy channels into the solar corona}",
      journal = {\nat},
         year = 2012,
        month = jun,
       volume = {486},
       number = {7404},
        pages = {505-508},
          doi = {10.1038/nature11202},
       adsurl = {https://ui.adsabs.harvard.edu/abs/2012Natur.486..505W}
}

@ARTICLE{Wedemeyer2013,
       author = {{Wedemeyer}, Sven and {Scullion}, Eamon and {Rouppe van der Voort}, Luc and {Bosnjak}, Antonija and {Antolin}, Patrick},
        title = "{Are Giant Tornadoes the Legs of Solar Prominences?}",
      journal = {\apj},
         year = 2013,
        month = sep,
       volume = {774},
       number = {2},
          eid = {123},
        pages = {123},
          doi = {10.1088/0004-637X/774/2/123},
archivePrefix = {arXiv},
       eprint = {1306.2661},
 primaryClass = {astro-ph.SR},
       adsurl = {https://ui.adsabs.harvard.edu/abs/2013ApJ...774..123W}
}

@ARTICLE{Panesar2013,
       author = {{Panesar}, N.~K. and {Innes}, D.~E. and {Tiwari}, S.~K. and {Low}, B.~C.},
        title = "{A solar tornado triggered by flares?}",
      journal = {\aap},
         year = 2013,
        month = jan,
       volume = {549},
          eid = {A105},
        pages = {A105},
          doi = {10.1051/0004-6361/201220503},
archivePrefix = {arXiv},
       eprint = {1211.6569},
 primaryClass = {astro-ph.SR},
       adsurl = {https://ui.adsabs.harvard.edu/abs/2013A&A...549A.105P}
}

@ARTICLE{Wedemeyer2014,
       author = {{Wedemeyer}, Sven and {Steiner}, Oskar},
        title = "{On the plasma flow inside magnetic tornadoes on the Sun}",
      journal = {\pasj},
         year = 2014,
        month = dec,
       volume = {66},
          eid = {S10},
        pages = {S10},
          doi = {10.1093/pasj/psu086},
archivePrefix = {arXiv},
       eprint = {1406.7270},
 primaryClass = {astro-ph.SR},
       adsurl = {https://ui.adsabs.harvard.edu/abs/2014PASJ...66S..10W}
}

@ARTICLE{Wang2017,
       author = {{Wang}, Wensi and {Liu}, Rui and {Wang}, Yuming},
        title = "{Tornado-like Evolution of a Kink-unstable Solar Prominence}",
      journal = {\apj},
         year = 2017,
        month = jan,
       volume = {834},
       number = {1},
          eid = {38},
        pages = {38},
          doi = {10.3847/1538-4357/834/1/38},
archivePrefix = {arXiv},
       eprint = {1611.04667},
 primaryClass = {astro-ph.SR},
       adsurl = {https://ui.adsabs.harvard.edu/abs/2017ApJ...834...38W}
}

@ARTICLE{Tziotziou2018,
       author = {{Tziotziou}, K. and {Tsiropoula}, G. and {Kontogiannis}, I. and {Scullion}, E. and {Doyle}, J.~G.},
        title = "{A persistent quiet-Sun small-scale tornado. I. Characteristics and dynamics}",
      journal = {\aap},
         year = 2018,
        month = oct,
       volume = {618},
          eid = {A51},
        pages = {A51},
          doi = {10.1051/0004-6361/201833101},
       adsurl = {https://ui.adsabs.harvard.edu/abs/2018A&A...618A..51T}
}

@ARTICLE{Shelyag2011,
       author = {{Shelyag}, S. and {Fedun}, V. and {Keenan}, F.~P. and {Erd{\'e}lyi}, R. and {Mathioudakis}, M.},
        title = "{Photospheric magnetic vortex structures}",
      journal = {Annales Geophysicae},
         year = 2011,
        month = may,
       volume = {29},
       number = {5},
        pages = {883-887},
          doi = {10.5194/angeo-29-883-2011},
       adsurl = {https://ui.adsabs.harvard.edu/abs/2011AnGeo..29..883S}
}

@ARTICLE{Murawski2018,
       author = {{Murawski}, K. and {Kayshap}, P. and {Srivastava}, A.~K. and {Pascoe}, D.~J. and {Jel{\'\i}nek}, P. and {Ku{\'z}ma}, B. and {Fedun}, V.},
        title = "{Magnetic swirls and associated fast magnetoacoustic kink waves in a solar chromospheric flux tube}",
      journal = {\mnras},
         year = 2018,
        month = feb,
       volume = {474},
       number = {1},
        pages = {77-87},
          doi = {10.1093/mnras/stx2763},
archivePrefix = {arXiv},
       eprint = {1710.08179},
 primaryClass = {astro-ph.SR},
       adsurl = {https://ui.adsabs.harvard.edu/abs/2018MNRAS.474...77M}
}

@ARTICLE{Liu2019NatCo,
       author = {{Liu}, Jiajia and {Nelson}, Chris J. and {Snow}, Ben and {Wang}, Yuming and {Erd{\'e}lyi}, Robert},
        title = "{Evidence of ubiquitous Alfv{\'e}n pulses transporting energy from the photosphere to the upper chromosphere}",
      journal = {Nature Communications},
         year = 2019,
        month = aug,
       volume = {10},
          eid = {3504},
        pages = {3504},
          doi = {10.1038/s41467-019-11495-0},
       adsurl = {https://ui.adsabs.harvard.edu/abs/2019NatCo..10.3504L}
}

@ARTICLE{Yadav2020,
       author = {{Yadav}, Nitin and {Cameron}, R.~H. and {Solanki}, S.~K.},
        title = "{Simulations Show that Vortex Flows Could Heat the Chromosphere in Solar Plage}",
      journal = {\apjl},
         year = 2020,
        month = may,
       volume = {894},
       number = {2},
          eid = {L17},
        pages = {L17},
          doi = {10.3847/2041-8213/ab8dc5},
archivePrefix = {arXiv},
       eprint = {2004.13996},
 primaryClass = {astro-ph.SR},
       adsurl = {https://ui.adsabs.harvard.edu/abs/2020ApJ...894L..17Y}
}

@ARTICLE{Scalisi2021,
       author = {{Scalisi}, Joseph and {Oxley}, William and {Ruderman}, Michael S. and {Erd{\'e}lyi}, Robertus},
        title = "{Propagation of Torsional Alfv{\'e}n Pulses in Zero-beta Flux Tubes}",
      journal = {\apj},
         year = 2021,
        month = apr,
       volume = {911},
       number = {1},
          eid = {39},
        pages = {39},
          doi = {10.3847/1538-4357/abe8db},
       adsurl = {https://ui.adsabs.harvard.edu/abs/2021ApJ...911...39S}
}

@ARTICLE{Scalisi2023,
       author = {{Scalisi}, Joseph and {Ruderman}, Michael S. and {Erd{\'e}lyi}, Robertus},
        title = "{Generation of Vertical Flows by Torsional Alfv{\'e}n Pulses in Zero-beta Tubes with a Transitional Layer}",
      journal = {\apj},
         year = 2023,
        month = jul,
       volume = {951},
       number = {1},
          eid = {60},
        pages = {60},
          doi = {10.3847/1538-4357/acd9ae},
       adsurl = {https://ui.adsabs.harvard.edu/abs/2023ApJ...951...60S}
}

@ARTICLE{Breu2023,
       author = {{Breu}, C. and {Peter}, H. and {Cameron}, R. and {Solanki}, S.~K.},
        title = "{Swirls in the solar corona}",
      journal = {\aap},
         year = 2023,
        month = jul,
       volume = {675},
          eid = {A94},
        pages = {A94},
          doi = {10.1051/0004-6361/202245780},
archivePrefix = {arXiv},
       eprint = {2305.03769},
 primaryClass = {astro-ph.SR},
       adsurl = {https://ui.adsabs.harvard.edu/abs/2023A&A...675A..94B}
}

@ARTICLE{Tziotziou2023,
       author = {{Tziotziou}, K. and {Scullion}, E. and {Shelyag}, S. and {Steiner}, O. and {Khomenko}, E. and {Tsiropoula}, G. and {Canivete Cuissa}, J.~R. and {Wedemeyer}, S. and {Kontogiannis}, I. and {Yadav}, N. and {Kitiashvili}, I.~N. and {Skirvin}, S.~J. and {Dakanalis}, I. and {Kosovichev}, A.~G. and {Fedun}, V.},
        title = "{Vortex Motions in the Solar Atmosphere}",
      journal = {\ssr},
         year = 2023,
        month = feb,
       volume = {219},
       number = {1},
          eid = {1},
        pages = {1},
          doi = {10.1007/s11214-022-00946-8},
       adsurl = {https://ui.adsabs.harvard.edu/abs/2023SSRv..219....1T}
}

@ARTICLE{Finley2022,
       author = {{Finley}, A.~J. and {Brun}, A.~S. and {Carlsson}, M. and {Szydlarski}, M. and {Hansteen}, V. and {Shoda}, M.},
        title = "{Stirring the base of the solar wind: On heat transfer and vortex formation}",
      journal = {\aap},
         year = 2022,
        month = sep,
       volume = {665},
          eid = {A118},
        pages = {A118},
          doi = {10.1051/0004-6361/202243947},
archivePrefix = {arXiv},
       eprint = {2207.02878},
 primaryClass = {astro-ph.SR},
       adsurl = {https://ui.adsabs.harvard.edu/abs/2022A&A...665A.118F}
}

@ARTICLE{Battaglia2021,
       author = {{Battaglia}, Andrea Francesco and {Canivete Cuissa}, Jos{\'e} Roberto and {Calvo}, Flavio and {Bossart}, Aleksi Antoine and {Steiner}, Oskar},
        title = "{The Alfv{\'e}nic nature of chromospheric swirls}",
      journal = {\aap},
         year = 2021,
        month = may,
       volume = {649},
          eid = {A121},
        pages = {A121},
          doi = {10.1051/0004-6361/202040110},
archivePrefix = {arXiv},
       eprint = {2103.07366},
 primaryClass = {astro-ph.SR},
       adsurl = {https://ui.adsabs.harvard.edu/abs/2021A&A...649A.121B}
}

@ARTICLE{Kuniyoshi2023,
       author = {{Kuniyoshi}, Hidetaka and {Shoda}, Munehito and {Iijima}, Haruhisa and {Yokoyama}, Takaaki},
        title = "{Magnetic Tornado Properties: A Substantial Contribution to the Solar Coronal Heating via Efficient Energy Transfer}",
      journal = {\apj},
         year = 2023,
        month = may,
       volume = {949},
       number = {1},
          eid = {8},
        pages = {8},
          doi = {10.3847/1538-4357/accbb8},
archivePrefix = {arXiv},
       eprint = {2304.03010},
 primaryClass = {astro-ph.SR},
       adsurl = {https://ui.adsabs.harvard.edu/abs/2023ApJ...949....8K}
}

@ARTICLE{Beckers1963,
       author = {{Beckers}, Jacques M.},
        title = "{Study of the Undisturbed Chromosphere from Ha-DISK Filtergrams, with Particular Reference to the Indentification of Spicules.}",
      journal = {\apj},
         year = 1963,
        month = oct,
       volume = {138},
        pages = {648},
          doi = {10.1086/147676},
       adsurl = {https://ui.adsabs.harvard.edu/abs/1963ApJ...138..648B}
}

@ARTICLE{Sterling2000,
       author = {{Sterling}, Alphonse C.},
        title = "{Solar Spicules: A Review of Recent Models and Targets for Future Observations - (Invited Review)}",
      journal = {\solphys},
         year = 2000,
        month = sep,
       volume = {196},
       number = {1},
        pages = {79-111},
          doi = {10.1023/A:1005213923962},
       adsurl = {https://ui.adsabs.harvard.edu/abs/2000SoPh..196...79S}
}

@ARTICLE{Zaqarashvili2009,
       author = {{Zaqarashvili}, T.~V. and {Erd{\'e}lyi}, R.},
        title = "{Oscillations and Waves in Solar Spicules}",
      journal = {\ssr},
         year = 2009,
        month = dec,
       volume = {149},
       number = {1-4},
        pages = {355-388},
          doi = {10.1007/s11214-009-9549-y},
archivePrefix = {arXiv},
       eprint = {0906.1783},
 primaryClass = {astro-ph.SR},
       adsurl = {https://ui.adsabs.harvard.edu/abs/2009SSRv..149..355Z}
}

@ARTICLE{Tsiropoula2012,
       author = {{Tsiropoula}, G. and {Tziotziou}, K. and {Kontogiannis}, I. and {Madjarska}, M.~S. and {Doyle}, J.~G. and {Suematsu}, Y.},
        title = "{Solar Fine-Scale Structures. I. Spicules and Other Small-Scale, Jet-Like Events at the Chromospheric Level: Observations and Physical Parameters}",
      journal = {\ssr},
         year = 2012,
        month = sep,
       volume = {169},
       number = {1-4},
        pages = {181-244},
          doi = {10.1007/s11214-012-9920-2},
archivePrefix = {arXiv},
       eprint = {1207.3956},
 primaryClass = {astro-ph.SR},
       adsurl = {https://ui.adsabs.harvard.edu/abs/2012SSRv..169..181T}
}

@ARTICLE{Shibata1982,
       author = {{Shibata}, K. and {Suematsu}, Y.},
        title = "{Why are spicules absent over plages and long under coronal holes?}",
      journal = {\solphys},
         year = 1982,
        month = jun,
       volume = {78},
       number = {2},
        pages = {333-345},
          doi = {10.1007/BF00151612},
       adsurl = {https://ui.adsabs.harvard.edu/abs/1982SoPh...78..333S}
}

@ARTICLE{Iijima2015,
       author = {{Iijima}, H. and {Yokoyama}, T.},
        title = "{Effect of Coronal Temperature on the Scale of Solar Chromospheric Jets}",
      journal = {\apjl},
         year = 2015,
        month = oct,
       volume = {812},
       number = {2},
          eid = {L30},
        pages = {L30},
          doi = {10.1088/2041-8205/812/2/L30},
archivePrefix = {arXiv},
       eprint = {1509.06677},
 primaryClass = {astro-ph.SR},
       adsurl = {https://ui.adsabs.harvard.edu/abs/2015ApJ...812L..30I}
}

@ARTICLE{Hollweg1982,
       author = {{Hollweg}, J.~V. and {Jackson}, S. and {Galloway}, D.},
        title = "{Alfven Waves in the Solar Atmospheres - Part Three - Nonlinear Waves on Open Flux Tubes}",
      journal = {\solphys},
         year = 1982,
        month = jan,
       volume = {75},
       number = {1-2},
        pages = {35-61},
          doi = {10.1007/BF00153458},
       adsurl = {https://ui.adsabs.harvard.edu/abs/1982SoPh...75...35H}
}

@ARTICLE{dePontieu1998,
       author = {{de Pontieu}, B. and {Haerendel}, G.},
        title = "{Weakly damped Alfven waves as drivers for spicules}",
      journal = {\aap},
         year = 1998,
        month = oct,
       volume = {338},
        pages = {729-736},
       adsurl = {https://ui.adsabs.harvard.edu/abs/1998A&A...338..729D}
}

@ARTICLE{James2002,
       author = {{James}, S.~P. and {Erd{\'e}lyi}, R.},
        title = "{Spicule formation by ion-neutral damping}",
      journal = {\aap},
         year = 2002,
        month = oct,
       volume = {393},
        pages = {L11-L14},
          doi = {10.1051/0004-6361:20021126},
       adsurl = {https://ui.adsabs.harvard.edu/abs/2002A&A...393L..11J}
}

@ARTICLE{Yokoyama1995,
       author = {{Yokoyama}, Takaaki and {Shibata}, Kazunari},
        title = "{Magnetic reconnection as the origin of X-ray jets and H{\ensuremath{\alpha}} surges on the Sun}",
      journal = {\nat},
         year = 1995,
        month = may,
       volume = {375},
       number = {6526},
        pages = {42-44},
          doi = {10.1038/375042a0},
       adsurl = {https://ui.adsabs.harvard.edu/abs/1995Natur.375...42Y}
}

@ARTICLE{Singh2011,
       author = {{Singh}, K.~A.~P. and {Shibata}, K. and {Nishizuka}, N. and {Isobe}, H.},
        title = "{Chromospheric anemone jets and magnetic reconnection in partially ionized solar atmosphere}",
      journal = {Physics of Plasmas},
         year = 2011,
        month = nov,
       volume = {18},
       number = {11},
        pages = {111210-111210},
          doi = {10.1063/1.3655444},
       adsurl = {https://ui.adsabs.harvard.edu/abs/2011PhPl...18k1210S}
}

@ARTICLE{Archontis2010,
       author = {{Archontis}, V. and {Tsinganos}, K. and {Gontikakis}, C.},
        title = "{Recurrent solar jets in active regions}",
      journal = {\aap},
         year = 2010,
        month = mar,
       volume = {512},
          eid = {L2},
        pages = {L2},
          doi = {10.1051/0004-6361/200913752},
archivePrefix = {arXiv},
       eprint = {1003.2349},
 primaryClass = {astro-ph.SR},
       adsurl = {https://ui.adsabs.harvard.edu/abs/2010A&A...512L...2A}
}

@ARTICLE{Mart2017,
       author = {{Mart{\'\i}nez-Sykora}, J. and {De Pontieu}, B. and {Hansteen}, V.~H. and {Rouppe van der Voort}, L. and {Carlsson}, M. and {Pereira}, T.~M.~D.},
        title = "{On the generation of solar spicules and Alfv{\'e}nic waves}",
      journal = {Science},
         year = 2017,
        month = jun,
       volume = {356},
       number = {6344},
        pages = {1269-1272},
          doi = {10.1126/science.aah5412},
archivePrefix = {arXiv},
       eprint = {1710.07559},
 primaryClass = {astro-ph.SR},
       adsurl = {https://ui.adsabs.harvard.edu/abs/2017Sci...356.1269M}
}

@ARTICLE{DePontieu2014,
       author = {{De Pontieu}, B. and {Rouppe van der Voort}, L. and {McIntosh}, S.~W. and {Pereira}, T.~M.~D. and {Carlsson}, M. and {Hansteen}, V. and {Skogsrud}, H. and {Lemen}, J. and {Title}, A. and {Boerner}, P. and {Hurlburt}, N. and {Tarbell}, T.~D. and {Wuelser}, J.~P. and {De Luca}, E.~E. and {Golub}, L. and {McKillop}, S. and {Reeves}, K. and {Saar}, S. and {Testa}, P. and {Tian}, H. and {Kankelborg}, C. and {Jaeggli}, S. and {Kleint}, L. and {Martinez-Sykora}, J.},
        title = "{On the prevalence of small-scale twist in the solar chromosphere and transition region}",
      journal = {Science},
         year = 2014,
        month = oct,
       volume = {346},
       number = {6207},
          eid = {1255732},
        pages = {1255732},
          doi = {10.1126/science.1255732},
archivePrefix = {arXiv},
       eprint = {1410.6862},
 primaryClass = {astro-ph.SR},
       adsurl = {https://ui.adsabs.harvard.edu/abs/2014Sci...346D.315D}
}

@ARTICLE{DePontieu2012,
       author = {{De Pontieu}, B. and {Carlsson}, M. and {Rouppe van der Voort}, L.~H.~M. and {Rutten}, R.~J. and {Hansteen}, V.~H. and {Watanabe}, H.},
        title = "{Ubiquitous Torsional Motions in Type II Spicules}",
      journal = {\apjl},
         year = 2012,
        month = jun,
       volume = {752},
       number = {1},
          eid = {L12},
        pages = {L12},
          doi = {10.1088/2041-8205/752/1/L12},
archivePrefix = {arXiv},
       eprint = {1205.5006},
 primaryClass = {astro-ph.SR},
       adsurl = {https://ui.adsabs.harvard.edu/abs/2012ApJ...752L..12D}
}

@ARTICLE{Kitiashvili2013,
       author = {{Kitiashvili}, I.~N. and {Kosovichev}, A.~G. and {Lele}, S.~K. and {Mansour}, N.~N. and {Wray}, A.~A.},
        title = "{Ubiquitous Solar Eruptions Driven by Magnetized Vortex Tubes}",
      journal = {\apj},
         year = 2013,
        month = jun,
       volume = {770},
       number = {1},
          eid = {37},
        pages = {37},
          doi = {10.1088/0004-637X/770/1/37},
archivePrefix = {arXiv},
       eprint = {1301.0018},
 primaryClass = {astro-ph.SR},
       adsurl = {https://ui.adsabs.harvard.edu/abs/2013ApJ...770...37K}
}

@ARTICLE{Iijima2017,
       author = {{Iijima}, H. and {Yokoyama}, T.},
        title = "{A Three-dimensional Magnetohydrodynamic Simulation of the Formation of Solar Chromospheric Jets with Twisted Magnetic Field Lines}",
      journal = {\apj},
         year = 2017,
        month = oct,
       volume = {848},
       number = {1},
          eid = {38},
        pages = {38},
          doi = {10.3847/1538-4357/aa8ad1},
archivePrefix = {arXiv},
       eprint = {1709.01522},
 primaryClass = {astro-ph.SR},
       adsurl = {https://ui.adsabs.harvard.edu/abs/2017ApJ...848...38I}
}

@ARTICLE{Snow2018,
       author = {{Snow}, B. and {Fedun}, V. and {Gent}, F.~A. and {Verth}, G. and {Erd{\'e}lyi}, R.},
        title = "{Magnetic Shocks and Substructures Excited by Torsional Alfv{\'e}n Wave Interactions in Merging Expanding Flux Tubes}",
      journal = {\apj},
         year = 2018,
        month = apr,
       volume = {857},
       number = {2},
          eid = {125},
        pages = {125},
          doi = {10.3847/1538-4357/aab7f7},
archivePrefix = {arXiv},
       eprint = {1803.06112},
 primaryClass = {astro-ph.SR},
       adsurl = {https://ui.adsabs.harvard.edu/abs/2018ApJ...857..125S}
}

@ARTICLE{Rempel2017,
       author = {{Rempel}, M.},
        title = "{Extension of the MURaM Radiative MHD Code for Coronal Simulations}",
      journal = {\apj},
         year = 2017,
        month = jan,
       volume = {834},
       number = {1},
          eid = {10},
        pages = {10},
          doi = {10.3847/1538-4357/834/1/10},
archivePrefix = {arXiv},
       eprint = {1609.09818},
 primaryClass = {astro-ph.SR},
       adsurl = {https://ui.adsabs.harvard.edu/abs/2017ApJ...834...10R}
}

@ARTICLE{Cheng2025,
       author = {{Cheng}, Guanchong and {Ni}, Lei and {Cameron}, Robert and {Peter}, Hardi and {Chen}, Yajie and {Lin}, Jun},
        title = "{Three-dimensional magnetic reconnection mediated with plasmoids and the resulting multithermal emissions in the cool atmosphere of the Sun}",
      journal = {\aap},
         year = 2025,
        month = sep,
       volume = {701},
          eid = {A245},
        pages = {A245},
          doi = {10.1051/0004-6361/202554204},
archivePrefix = {arXiv},
       eprint = {2508.11013},
 primaryClass = {astro-ph.SR},
       adsurl = {https://ui.adsabs.harvard.edu/abs/2025A&A...701A.245C}
}

@ARTICLE{Gudiksen2011,
       author = {{Gudiksen}, B.~V. and {Carlsson}, M. and {Hansteen}, V.~H. and {Hayek}, W. and {Leenaarts}, J. and {Mart{\'\i}nez-Sykora}, J.},
        title = "{The stellar atmosphere simulation code Bifrost. Code description and validation}",
      journal = {\aap},
         year = 2011,
        month = jul,
       volume = {531},
          eid = {A154},
        pages = {A154},
          doi = {10.1051/0004-6361/201116520},
archivePrefix = {arXiv},
       eprint = {1105.6306},
 primaryClass = {astro-ph.SR},
       adsurl = {https://ui.adsabs.harvard.edu/abs/2011A&A...531A.154G}
}

@ARTICLE{Welsch2015,
       author = {{Welsch}, Brian T.},
        title = "{The photospheric Poynting flux and coronal heating}",
      journal = {\pasj},
         year = 2015,
        month = apr,
       volume = {67},
       number = {2},
          eid = {18},
        pages = {18},
          doi = {10.1093/pasj/psu151},
archivePrefix = {arXiv},
       eprint = {1402.4794},
 primaryClass = {astro-ph.SR},
       adsurl = {https://ui.adsabs.harvard.edu/abs/2015PASJ...67...18W}
}

@ARTICLE{Mumford2015,
       author = {{Mumford}, S.~J. and {Fedun}, V. and {Erd{\'e}lyi}, R.},
        title = "{Generation of Magnetohydrodynamic Waves in Low Solar Atmospheric Flux Tubes by Photospheric Motions}",
      journal = {\apj},
         year = 2015,
        month = jan,
       volume = {799},
       number = {1},
          eid = {6},
        pages = {6},
          doi = {10.1088/0004-637X/799/1/6},
archivePrefix = {arXiv},
       eprint = {1305.7415},
 primaryClass = {astro-ph.SR},
       adsurl = {https://ui.adsabs.harvard.edu/abs/2015ApJ...799....6M}
}

@ARTICLE{Khomenko2012,
       author = {{Khomenko}, E. and {Cally}, P.~S.},
        title = "{Numerical Simulations of Conversion to Alfv{\'e}n Waves in Sunspots}",
      journal = {\apj},
         year = 2012,
        month = feb,
       volume = {746},
       number = {1},
          eid = {68},
        pages = {68},
          doi = {10.1088/0004-637X/746/1/68},
archivePrefix = {arXiv},
       eprint = {1111.2851},
 primaryClass = {astro-ph.SR},
       adsurl = {https://ui.adsabs.harvard.edu/abs/2012ApJ...746...68K}
}

@ARTICLE{Vigeesh2012,
       author = {{Vigeesh}, G. and {Fedun}, V. and {Hasan}, S.~S. and {Erd{\'e}lyi}, R.},
        title = "{Three-dimensional Simulations of Magnetohydrodynamic Waves in Magnetized Solar Atmosphere}",
      journal = {\apj},
         year = 2012,
        month = aug,
       volume = {755},
       number = {1},
          eid = {18},
        pages = {18},
          doi = {10.1088/0004-637X/755/1/18},
archivePrefix = {arXiv},
       eprint = {1109.6471},
 primaryClass = {astro-ph.SR},
       adsurl = {https://ui.adsabs.harvard.edu/abs/2012ApJ...755...18V}
}

@ARTICLE{Li2026,
       author = {{Li}, Jin and {Ni}, Lei and {Erd{\'e}lyi}, Robertus and {Lin}, Jun and {Yan}, Xiaoli and {Cheng}, Guanchong},
        title = "{Radiative MHD Studies of Solar Spicules: Driving Mechanisms and the Role of Background Magnetic Field}",
      journal = {Research in Astronomy and Astrophysics},
         year = 2026,
        month = sep,
       volume = {26},
       number = {9},
          eid = {095015},
        pages = {095015},
          doi = {10.1088/1674-4527/ae616d},
archivePrefix = {arXiv},
       eprint = {2605.24374},
 primaryClass = {astro-ph.SR},
       adsurl = {https://ui.adsabs.harvard.edu/abs/2026RAA....26i5015L}
}

@ARTICLE{Ni2026,
       author = {{Ni}, Lei and {Lin}, Jun and {Samanta}, Tanmoy and {Cheng}, Guanchong and {Wang}, Yifu and {Erd{\'e}lyi}, Robert},
        title = "{Energy and Mass Transport Associated with Impulsive Spicular Flows in Solar Coronal Holes}",
      journal = {\apjl},
         year = 2026,
        month = mar,
       volume = {1000},
       number = {1},
          eid = {L24},
        pages = {L24},
          doi = {10.3847/2041-8213/ae4964},
archivePrefix = {arXiv},
       eprint = {2602.19953},
 primaryClass = {astro-ph.SR},
       adsurl = {https://ui.adsabs.harvard.edu/abs/2026ApJ..1000L..24N}
}

@ARTICLE{Rempel2014,
       author = {{Rempel}, M.},
        title = "{Numerical Simulations of Quiet Sun Magnetism: On the Contribution from a Small-scale Dynamo}",
      journal = {\apj},
         year = 2014,
        month = jul,
       volume = {789},
       number = {2},
          eid = {132},
        pages = {132},
          doi = {10.1088/0004-637X/789/2/132},
archivePrefix = {arXiv},
       eprint = {1405.6814},
 primaryClass = {astro-ph.SR},
       adsurl = {https://ui.adsabs.harvard.edu/abs/2014ApJ...789..132R}
}

@ARTICLE{Chandra2025,
       author = {{Chandra}, Sanghita and {Cameron}, Robert and {Przybylski}, Damien and {Solanki}, Sami K. and {Ondratschek}, Patrick and {Danilovic}, Sanja},
        title = "{Probing chromospheric fine structures with a H{\ensuremath{\alpha}} proxy using MURaM-ChE}",
      journal = {\aap},
         year = 2025,
        month = sep,
       volume = {701},
          eid = {A294},
        pages = {A294},
          doi = {10.1051/0004-6361/202555646},
archivePrefix = {arXiv},
       eprint = {2508.04387},
 primaryClass = {astro-ph.SR},
       adsurl = {https://ui.adsabs.harvard.edu/abs/2025A&A...701A.294C}
}

@ARTICLE{Liu2019b,
       author = {{Liu}, Jiajia and {Nelson}, Chris J. and {Erd{\'e}lyi}, Robertus},
        title = "{Automated Swirl Detection Algorithm (ASDA) and Its Application to Simulation and Observational Data}",
      journal = {\apj},
         year = 2019,
        month = feb,
       volume = {872},
       number = {1},
          eid = {22},
        pages = {22},
          doi = {10.3847/1538-4357/aabd34},
archivePrefix = {arXiv},
       eprint = {1804.02931},
 primaryClass = {astro-ph.SR},
       adsurl = {https://ui.adsabs.harvard.edu/abs/2019ApJ...872...22L}
}

@ARTICLE{Yadav2021,
       author = {{Yadav}, N. and {Cameron}, R.~H. and {Solanki}, S.~K.},
        title = "{Vortex flow properties in simulations of solar plage region: Evidence for their role in chromospheric heating}",
      journal = {\aap},
         year = 2021,
        month = jan,
       volume = {645},
          eid = {A3},
        pages = {A3},
          doi = {10.1051/0004-6361/202038965},
archivePrefix = {arXiv},
       eprint = {2010.14971},
 primaryClass = {astro-ph.SR},
       adsurl = {https://ui.adsabs.harvard.edu/abs/2021A&A...645A...3Y}
}

@ARTICLE{Przybylski2022,
       author = {{Przybylski}, D. and {Cameron}, R. and {Solanki}, S.~K. and {Rempel}, M. and {Leenaarts}, J. and {Anusha}, L.~S. and {Witzke}, V. and {Shapiro}, A.~I.},
        title = "{Chromospheric extension of the MURaM code}",
      journal = {\aap},
         year = 2022,
        month = aug,
       volume = {664},
          eid = {A91},
        pages = {A91},
          doi = {10.1051/0004-6361/202141230},
archivePrefix = {arXiv},
       eprint = {2204.03126},
 primaryClass = {astro-ph.SR},
       adsurl = {https://ui.adsabs.harvard.edu/abs/2022A&A...664A..91P}
}

@ARTICLE{Yang2013,
       author = {{Yang}, Liping and {He}, Jiansen and {Peter}, Hardi and {Tu}, Chuanyi and {Zhang}, Lei and {Feng}, Xueshang and {Zhang}, Shaohua},
        title = "{Numerical Simulations of Chromospheric Anemone Jets Associated with Moving Magnetic Features}",
      journal = {\apj},
         year = 2013,
        month = nov,
       volume = {777},
       number = {1},
          eid = {16},
        pages = {16},
          doi = {10.1088/0004-637X/777/1/16},
       adsurl = {https://ui.adsabs.harvard.edu/abs/2013ApJ...777...16Y}
}

@ARTICLE{Chen2026,
       author = {{Chen},Yajie and {Peter}, Hardi and {Przybylski}, Damien and {Chitta}, Lakshmi Pradeep and {Mandal}, Sudip},
        title = "{A self-consistent 3D magnetohydrodynamic model producing a solar blowout jet}",
      journal = {\aap},
         year = 2026,
        month = feb,
       volume = {706},
          eid = {A141},
        pages = {A141},
          doi = {10.1051/0004-6361/202556660},
archivePrefix = {arXiv},
       eprint = {2512.21728},
 primaryClass = {astro-ph.SR},
       adsurl = {https://ui.adsabs.harvard.edu/abs/2026A&A...706A.141C}
}

@ARTICLE{Attie2009,
       author = {{Attie}, R. and {Innes}, D.~E. and {Potts}, H.~E.},
        title = "{Evidence of photospheric vortex flows at supergranular junctions observed by FG/SOT (Hinode)}",
      journal = {\aap},
         year = 2009,
        month = jan,
       volume = {493},
       number = {2},
        pages = {L13-L16},
          doi = {10.1051/0004-6361:200811258},
archivePrefix = {arXiv},
       eprint = {0811.3445},
 primaryClass = {astro-ph},
       adsurl = {https://ui.adsabs.harvard.edu/abs/2009A&A...493L..13A}
}

@ARTICLE{Balmaceda2010,
       author = {{Balmaceda}, L. and {Vargas Dom{\'\i}nguez}, S. and {Palacios}, J. and {Cabello}, I. and {Domingo}, V.},
        title = "{Evidence of small-scale magnetic concentrations dragged by vortex motion of solar photospheric plasma}",
      journal = {\aap},
         year = 2010,
        month = apr,
       volume = {513},
          eid = {L6},
        pages = {L6},
          doi = {10.1051/0004-6361/200913584},
archivePrefix = {arXiv},
       eprint = {1004.1185},
 primaryClass = {astro-ph.SR},
       adsurl = {https://ui.adsabs.harvard.edu/abs/2010A&A...513L...6B}
}

@ARTICLE{Giagkiozis2018,
       author = {{Giagkiozis}, Ioannis and {Fedun}, Viktor and {Scullion}, Eamon and {Jess}, David B. and {Verth}, Gary},
        title = "{Vortex Flows in the Solar Atmosphere: Automated Identification and Statistical Analysis}",
      journal = {\apj},
         year = 2018,
        month = dec,
       volume = {869},
       number = {2},
          eid = {169},
        pages = {169},
          doi = {10.3847/1538-4357/aaf797},
archivePrefix = {arXiv},
       eprint = {1706.05428},
 primaryClass = {astro-ph.SR},
       adsurl = {https://ui.adsabs.harvard.edu/abs/2018ApJ...869..169G}
}

@ARTICLE{Dey2022,
       author = {{Dey}, Sahel and {Chatterjee}, Piyali and {O.~V.~S.~N.}, Murthy and {Kors{\'o}s}, Marianna B. and {Liu}, Jiajia and {Nelson}, Christopher J. and {Erd{\'e}lyi}, Robertus},
        title = "{Polymeric jets throw light on the origin and nature of the forest of solar spicules}",
      journal = {Nature Physics},
         year = 2022,
        month = mar,
       volume = {18},
       number = {5},
        pages = {595-600},
          doi = {10.1038/s41567-022-01522-1},
archivePrefix = {arXiv},
       eprint = {2211.04493},
 primaryClass = {astro-ph.SR},
       adsurl = {https://ui.adsabs.harvard.edu/abs/2022NatPh..18..595D}
}

@ARTICLE{Srivastava2025,
       author = {{Srivastava}, Sankalp and {Chatterjee}, Piyali and {Dey}, Sahel and {Erd{\'e}lyi}, Robertus},
        title = "{The Relation between Solar Spicules and Magnetohydrodynamic Shocks}",
      journal = {\apj},
         year = 2025,
        month = aug,
       volume = {989},
       number = {1},
          eid = {39},
        pages = {39},
          doi = {10.3847/1538-4357/ade9b4},
archivePrefix = {arXiv},
       eprint = {2506.21517},
 primaryClass = {astro-ph.SR},
       adsurl = {https://ui.adsabs.harvard.edu/abs/2025ApJ...989...39S}
}

@ARTICLE{Mumford2015MN,
       author = {{Mumford}, S.~J. and {Erd{\'e}lyi}, R.},
        title = "{Photospheric logarithmic velocity spirals as MHD wave generation mechanisms}",
      journal = {\mnras},
         year = 2015,
        month = may,
       volume = {449},
       number = {2},
        pages = {1679-1685},
          doi = {10.1093/mnras/stv365},
archivePrefix = {arXiv},
       eprint = {1501.01871},
 primaryClass = {astro-ph.SR},
       adsurl = {https://ui.adsabs.harvard.edu/abs/2015MNRAS.449.1679M}
}

@ARTICLE{Leonard2018,
       author = {{Leonard}, A.~J. and {Mumford}, S.~J. and {Fedun}, V. and {Erd{\'e}lyi}, R.},
        title = "{Varying driver velocity fields in photospheric MHD wave simulations}",
      journal = {\mnras},
         year = 2018,
        month = oct,
       volume = {480},
       number = {2},
        pages = {2839-2845},
          doi = {10.1093/mnras/sty1981},
archivePrefix = {arXiv},
       eprint = {1807.10049},
 primaryClass = {astro-ph.SR},
       adsurl = {https://ui.adsabs.harvard.edu/abs/2018MNRAS.480.2839L}
}

@ARTICLE{Oxley2020,
       author = {{Oxley}, William and {Scalisi}, Joseph and {Ruderman}, Michael S. and {Erd{\'e}lyi}, R{\'o}bert},
        title = "{Formation of Chromospheric Spicules in Magnetic Bright Points: An Analytical Approach Using Cartesian Slab Geometry}",
      journal = {\apj},
         year = 2020,
        month = dec,
       volume = {905},
       number = {2},
          eid = {168},
        pages = {168},
          doi = {10.3847/1538-4357/abcafe},
       adsurl = {https://ui.adsabs.harvard.edu/abs/2020ApJ...905..168O}
}

@ARTICLE{Magara2003,
       author = {{Magara}, T. and {Longcope}, D.~W.},
        title = "{Injection of Magnetic Energy and Magnetic Helicity into the Solar Atmosphere by an Emerging Magnetic Flux Tube}",
      journal = {\apj},
         year = 2003,
        month = mar,
       volume = {586},
       number = {1},
        pages = {630-649},
          doi = {10.1086/367611},
       adsurl = {https://ui.adsabs.harvard.edu/abs/2003ApJ...586..630M}
}

@ARTICLE{Lee2026,
       author = {{Lee}, Jeongwoo and {Lim}, Eun-Kyung and {Hansteen}, Viggo},
        title = "{Spatial and Dynamical Relations between Spicules and Network Bright Points}",
      journal = {\apj},
         year = 2026,
        month = feb,
       volume = {997},
       number = {2},
          eid = {170},
        pages = {170},
          doi = {10.3847/1538-4357/ae2747},
archivePrefix = {arXiv},
       eprint = {2512.07681},
 primaryClass = {astro-ph.SR},
       adsurl = {https://ui.adsabs.harvard.edu/abs/2026ApJ...997..170L}
}

@ARTICLE{Chen2017,
       author = {{Chen}, Jie and {Su}, Jiangtao and {Deng}, Yuanyong and {Priest}, E.~R.},
        title = "{A Complex Solar Coronal Jet with Two Phases}",
      journal = {\apj},
         year = 2017,
        month = may,
       volume = {840},
       number = {1},
          eid = {54},
        pages = {54},
          doi = {10.3847/1538-4357/aa6c59},
archivePrefix = {arXiv},
       eprint = {1704.02072},
 primaryClass = {astro-ph.SR},
       adsurl = {https://ui.adsabs.harvard.edu/abs/2017ApJ...840...54C}
}

@ARTICLE{Cherry2026,
       author = {{Cherry}, George and {Gudiksen}, Boris and {Finley}, Adam J. and {Noraz}, Quentin},
        title = "{Decomposing wave activity in the solar atmosphere: Shocks, jets, and swirls in the quiet Sun}",
      journal = {\aap},
         year = 2026,
        month = jan,
       volume = {705},
          eid = {A42},
        pages = {A42},
          doi = {10.1051/0004-6361/202557226},
archivePrefix = {arXiv},
       eprint = {2509.11179},
 primaryClass = {astro-ph.SR},
       adsurl = {https://ui.adsabs.harvard.edu/abs/2026A&A...705A..42C}
}

@article{Pereira2012,
  author    = {{Pereira}, T. M. D. and {Pontieu}, B. D. and {Carlsson}, M.},
  title     = {Quantifying Spicules},
  journal   = {\apj},
  year      = {2012},
  volume    = {759},
  number    = {1},
  pages     = {1093--1094},
  doi       = {10.1088/0004-637X/759/1/1093},
  adsurl    = {https://ui.adsabs.harvard.edu/abs/2012ApJ...759L..38P}
}

@ARTICLE{2014Sci...346A.315T,
	author = {{Tian}, H and {DeLuca}, E.E. and {Cranmer}, S.R. and {De Pontieu}, B. and {Peter}, H. and {Martinez-Sykora}, J and {Golub}, L. and {McKillop}, S. and {Reeves}, K.K. and {Miralles}, M.P. and {McCauley}, P. },
	title = "{Prevalence of small-scale jets from the networks of the solar transition region and chromosphere}",
	journal = {Science},
	year = 2014,
	month = Oct,
	volume = {346},
	number = {6207},
	pages = {1255711},
	doi = {10.1126/science.1255711},
	primaryClass = {astro-ph},
	adsurl = {https://www.science.org/doi/10.1126/science.1255711}
}

@ARTICLE{2000SoPh..196...79S,
	author = {{Sterling}, Alphonse C.},
	title = "{Solar Spicules: A Review of Recent Models and Targets for Future Observations - (Invited Review) }",
	journal = {Sol. Phys.},
	year = 2000,
	month = Sep,
	volume = {196},
	pages = {79-111},
	doi = {10.1023/A:1005213923962},
	adsurl = {https://link.springer.com/article/10.1023/A:1005213923962}
}

@ARTICLE{2012SSRv..169..181T,
	author = {{Tsiropoula}, G. and {Tziotziou}, K. and {Kontogiannis}, I. and {Madjarska}, M. S. and {Doyle}, J. G. and {Suematsu}, Y. },
	title = "{Solar Fine-Scale Structures. I. Spicules and Other Small-Scale, Jet-Like Events at the Chromospheric Level: Observations and Physical Parameters }",
	journal = {Space Science Reviews},
	year =2012,
	month = Sep,
	volume = {169},
	pages = {181-244},
	doi = {10.1007/s11214-012-9920-2},
	adsurl = {https://link.springer.com/article/10.1007/s11214-012-9920-2}
}

@ARTICLE{2021RSPSA.47700217S,
	author = {{Shen}, Yuandeng},
	title = "{Observation and modelling of solar jets }",
	journal = {Proceedings of the Royal Society of London Series A},
	year = 2021,
	month = Feb,
	volume = {477},
	pages = {40},
	doi = {10.1098/rspa.2020.0217},
	adsurl = {https://royalsocietypublishing.org/doi/pdf/10.1098/rspa.2020.0217}
}

@ARTICLE{2019STP.....5b..58H,
       author = {{Huang}, Zhenghua and {Li}, Bo and {Xia}, Lidong},
        title = "{Observations of small-scale energetic events in the solar transition region: explosive events, UV bursts, and network jets}",
      journal = {Solar-Terrestrial Physics},
         year = 2019,
        month = jun,
       volume = {5},
       number = {2},
        pages = {58-68},
          doi = {10.12737/stp-52201909},
archivePrefix = {arXiv},
       eprint = {1904.06688},
 primaryClass = {astro-ph.SR},
       adsurl = {https://ui.adsabs.harvard.edu/abs/2019STP.....5b..58H}
}

@ARTICLE{2022A&A...657A.118Q,
       author = {{Qi}, Youqian and {Huang}, Zhenghua and {Xia}, Lidong and {Fu}, Hui and {Guo}, Mingzhe and {Hou}, Zhenyong and {Liu}, Weixin and {Sun}, Mingzhe and {Liu}, Dayang},
        title = "{Statistical properties of H{\ensuremath{\alpha}} jets in the polar coronal hole and their implications in coronal activities}",
      journal = {\aap},
         year = 2022,
        month = jan,
       volume = {657},
          eid = {A118},
        pages = {A118},
          doi = {10.1051/0004-6361/202141401},
archivePrefix = {arXiv},
       eprint = {2110.15543},
 primaryClass = {astro-ph.SR},
       adsurl = {https://ui.adsabs.harvard.edu/abs/2022A&A...657A.118Q}
}

@ARTICLE{2022A&A...668A.118C,
       author = {{Canivete Cuissa}, J.~R. and {Steiner}, O.},
        title = "{Innovative and automated method for vortex identification. I. Description of the SWIRL algorithm}",
      journal = {\aap},
         year = 2022,
        month = dec,
       volume = {668},
          eid = {A118},
        pages = {A118},
          doi = {10.1051/0004-6361/202243740},
archivePrefix = {arXiv},
       eprint = {2210.05223},
 primaryClass = {astro-ph.SR},
       adsurl = {https://ui.adsabs.harvard.edu/abs/2022A&A...668A.118C}
}

@ARTICLE{Chen2025,
       author = {{Chen}, Yajie and {Peter}, Hardi and {Przybylski}, Damien and {Iijima}, Haruhisa and {Chitta}, Lakshmi Pradeep},
        title = "{Magnetic reconnection sustains the mass budget of the solar wind}",
      journal = {\aap},
         year = 2025,
        month = oct,
       volume = {702},
          eid = {L4},
        pages = {L4},
          doi = {10.1051/0004-6361/202556696},
archivePrefix = {arXiv},
       eprint = {2509.11692},
 primaryClass = {astro-ph.SR},
       adsurl = {https://ui.adsabs.harvard.edu/abs/2025A&A...702L...4C}
}

@ARTICLE{2013ApJ...771...20M,
       author = {{Moreno-Insertis}, F. and {Galsgaard}, K.},
        title = "{Plasma Jets and Eruptions in Solar Coronal Holes: A Three-dimensional Flux Emergence Experiment}",
      journal = {\apj},
         year = 2013,
        month = jul,
       volume = {771},
       number = {1},
          eid = {20},
        pages = {20},
          doi = {10.1088/0004-637X/771/1/20},
archivePrefix = {arXiv},
       eprint = {1305.2201},
 primaryClass = {astro-ph.SR},
       adsurl = {https://ui.adsabs.harvard.edu/abs/2013ApJ...771...20M}
}

@ARTICLE{Dey2026,
       author = {{Dey}, Sahel and {Chatterjee}, Piyali and {Erd{\'e}lyi}, Robertus},
        title = "{Solar Spicule Rotation Driven by Plasma Curtain-Vortex Interactions}",
      journal = {\apj},
         year = 2026,
        month = sep,
       volume = {1008},
       number = {1},
          eid = {132},
        pages = {132},
          doi = {10.3847/1538-4357/ae8999},
       adsurl = {https://ui.adsabs.harvard.edu/abs/2026ApJ..1008..132D}
}

\end{document}